\documentclass[letterpaper,12pt]{article}
\usepackage{array,epsfig,rotating,setspace,latexsym,amsmath,epsf,amssymb,bm}
\usepackage{cite,graphicx,authblk,multirow,color,caption,subcaption}
\usepackage[title]{appendix}

\newtheorem{Theo}{Theorem}

\newtheorem{Lem}{Lemma}
\newtheorem{Cor}{Corollary}

\def\lf{\left\lfloor}
\def\rf{\right\rfloor}
\newcommand{\indep}{\rotatebox[origin=c]{90}{$\models$}}
\allowdisplaybreaks

\title{Secure Degrees of Freedom of One-hop Wireless Networks with No Eavesdropper CSIT\thanks{This work was supported by NSF Grants CNS 13-14733, CCF 14-22111 and CCF 14-22129, and presented in part at CISS 2013 and IEEE ISIT 2015.}}

\author{Pritam Mukherjee \qquad Jianwei Xie \qquad Sennur Ulukus\\
\normalsize Department of Electrical and Computer Engineering\\
\normalsize University of Maryland, College Park, MD 20742 \\
\normalsize {\it pritamm@umd.edu} \qquad {\it xiejw@umd.edu} \qquad {\it ulukus@umd.edu}}

\begin{document}
\maketitle

\vspace{-0.6cm}

\begin{abstract}
We consider three channel models: the wiretap channel with $M$ helpers, the $K$-user multiple access wiretap channel, and the $K$-user interference channel with an external eavesdropper, when no eavesdropper's channel state information (CSI) is available at the transmitters. In each case, we establish the optimal sum secure degrees of freedom (s.d.o.f.) by providing achievable schemes and matching converses. We show that the unavailability of the eavesdropper's CSIT does not reduce the s.d.o.f.~of the wiretap channel with helpers. However, there is loss in s.d.o.f.~for both the multiple access wiretap channel and the interference channel with an external eavesdropper. In particular, we show that in the absence of eavesdropper's CSIT, the $K$-user multiple access wiretap channel reduces to a wiretap channel with $(K-1)$ helpers from a sum s.d.o.f.~perspective, and the optimal sum s.d.o.f.~reduces from $\frac{K(K-1)}{K(K-1)+1}$ to $\frac{K-1}{K}$. For the interference channel with an external eavesdropper, the optimal sum s.d.o.f.~decreases from $\frac{K(K-1)}{2K-1}$ to $\frac{K-1}{2}$ in the absence of the eavesdropper's CSIT. Our results show that the lack of eavesdropper's CSIT does not have a significant impact on the optimal s.d.o.f.~for any of the three channel models, especially when the number of users is large. This implies that physical layer security can be made robust to the unavailability of eavesdropper CSIT at high signal to noise ratio (SNR) regimes by careful modification of the achievable schemes as demonstrated in this paper.
\end{abstract}

\section{Introduction}

The availability of channel state information at the transmitters (CSIT) plays a crucial role in securing wireless communication in the physical layer. In most practical scenarios, the channel gains are measured by the receivers and then fed back to the transmitters, which use the CSI to ensure security. A passive eavesdropper, however, cannot be expected to provide CSI for its channel. In this paper, we investigate how the unavailability of the eavesdropper's CSIT affects the optimal secure rates for three important channel models: the wiretap channel with helpers, the multiple access wiretap channel, and the interference channel with an external eavesdropper.

For each of these channel models, the secrecy capacity regions remain unknown, even with full eavesdropper CSIT. In the absence of exact capacity regions, we study the secure degrees of freedom (s.d.o.f.) of each channel model in the high signal-to-noise (SNR) regime. For the wiretap channel with $M$ helpers and full eavesdropper CSIT, references \cite{jianwei_ulukus_helper_2012,jianwei_ulukus_one_hop} determine the optimal s.d.o.f.~to be $\frac{M}{M+1}$. Further, references \cite{jianwei_ulukus_mac_2013,jianwei_ulukus_one_hop} determine the optimal sum s.d.o.f.~for the $K$-user multiple access wiretap channel with full eavesdropper CSIT to be $\frac{K(K-1)}{K(K-1)+1}$. For the interference channel with an external eavesdropper, the optimal sum s.d.o.f.~is shown to be $\frac{K(K-1)}{2K-1}$ in references \cite{jianwei_ulukus_interference_2013,jianwei_interference}, with full eavesdropper CSIT. In this paper, we focus on the case when no eavesdropper CSIT is available. We show that for the wiretap channel with $M$ helpers, an s.d.o.f.~of $\frac{M}{M+1}$ is achievable even without eavesdropper's CSIT; thus, there is no loss of s.d.o.f.~due to the unavailability of eavesdropper CSIT in this case. For the multiple access wiretap channel and the interference channel with an external eavesdropper, however, the optimal s.d.o.f.~decreases when there is no eavesdropper CSIT. In particular, without eavesdropper CSIT, the $K$-user multiple access wiretap channel reduces to a wiretap channel with $(K-1)$ helpers and the optimal sum s.d.o.f.~decreases from $\frac{K(K-1)}{K(K-1)+1}$ to $\frac{K-1}{K}$. For the interference channel with an external eavesdropper, the optimal sum s.d.o.f.~decreases from $\frac{K(K-1)}{2K-1}$ to $\frac{K-1}{2}$ in the absence of eavesdropper CSIT.

In order to establish the optimal sum s.d.o.f., we propose achievable schemes and provide matching converse proofs for each of these channel models. We note that any achievable scheme for the wiretap channel with $(K-1)$ helpers is also an achievable scheme for the $K$-user multiple access wiretap channel. Further, a converse for the $K$-user multiple access wiretap channel is an upper bound for the wiretap channel with $(K-1)$ helpers as well. Thus, we provide achievable schemes for the wiretap channel with helpers and a converse for the multiple access wiretap channel. We consider both fixed and fading channel gains. For the wiretap channel with helpers and the multiple access wiretap channel, we present schemes based on real interference alignment \cite{real_inter_align} and vector space alignment \cite{cadambe_jafar_interference} for fixed and fading channel gains, respectively. For the interference channel, our achievable schemes are based on asymptotic real alignment  \cite{real_inter_align,real_inter_align_exploit} and asymptotic vector space alignment \cite{cadambe_jafar_interference} for fixed and fading channel gains, respectively. For every channel model, we design our achievable schemes such that, the structure of the real alignment based scheme for the case of fixed channel gains is similar to that of the vector space alignment based scheme for the case of fading channels. Thus, our achievable schemes indicate a loose correspondence between the real and vector space alignment techniques.

For the interference channel with an external eavesdropper, as in \cite{jianwei_interference}, every transmitter sacrifices a part of its message space to transmit cooperative jamming signals in the form of artificial noise. However, instead of one artificial noise block as in \cite{jianwei_interference}, our scheme requires two noise blocks from each transmitter. The $2K$ noise blocks from the $K$ transmitters are then aligned at each legitimate receiver to occupy only $(K+1)$ \emph{block dimensions} out of the full space of $2K$ dimensions, thus, achieving $\frac{K-1}{2K}$ s.d.o.f.~per receiver. At the eavesdropper, however, the noise blocks do not align, and therefore, occupy the full space of $2K$ block dimensions, ensuring security of the message blocks. To the best of our knowledge, this is the first scheme in the literature which uses two noise blocks at each transmitter and aligns them in an optimal way to maximize the desired signal space at each legitimate receiver. An interesting aspect of our proposed schemes for the interference channel is that they provide confidentiality of the messages not only from the external eavesdropper but also from the unintended legitimate receivers. Thus, our schemes for both fixed and fading channel gains achieve the optimal sum s.d.o.f.~for the $K$-user interference channel with both \emph{confidential messages} and an external eavesdropper, with no eavesdropper CSIT.

To prove the converse, we combine techniques from \cite{jianwei_ulukus_one_hop,jianwei_interference} and \cite{aligned_image_sets_jafar}. We exploit a key result in \cite{aligned_image_sets_jafar} that the output entropy at a receiver whose CSIT is not available is at least as large as the output entropy at a receiver whose CSIT is available, even when the transmitters cooperate and transmit correlated signals. This result is similar in spirit to the \emph{least alignment lemma} in \cite{wiretap_helper_delayed}, where only \emph{linear} transmission strategies are considered. Intuitively, no alignment of signals is possible at the receiver whose CSIT is unavailable; therefore, the signals occupy the maximum possible space at that receiver. We combine this insight with the techniques of \cite{jianwei_ulukus_one_hop,jianwei_interference}. Specifically, we use discretized versions of the \emph{secrecy penalty lemma}, which quantifies the loss of rate due to the presence of an eavesdropper, and the \emph{role of a helper lemma}, which captures the trade-off, arising out of decodability constraints, between the message rate and the entropy of an independent helper signal. Together, these techniques enable us to establish the optimal sum s.d.o.f.~for the multiple access wiretap channel with no eavesdropper CSIT to be $\frac{K-1}{K}$ and the optimal sum s.d.o.f.~for the interference channel with an external eavesdropper and no eavesdropper CSIT to be $\frac{K-1}{2}$.

\subsection{Related Work} 

The secrecy capacity of the discrete memoryless wiretap channel is established in \cite{Wyner,csiszar}. The s.d.o.f.~of the single antenna Gaussian wiretap channel \cite{hellman}, and its variants \cite{liang, LiYatesTrappechapter, gopala, yates, PMukherjee_Ulukus2013} with different fading models and CSI availability conditions, is zero. In multi-user scenarios, however, positive s.d.o.f.~values can be achieved. Each transmitters may have independent messages of its own, as in multiple access wiretap channels introduced in \cite{tekin_yener_mac2008,tekin-yener-it2} and interference channels with confidential messages introduced in \cite{liu_maric_yates_BCCM_2008}, or may act as helpers as in \cite{lai_elgamal_relay, tang_helper}. While cooperative jamming strategies can improve the achievable rates \cite{tekin_yener_mac2008}, i.i.d.~Gaussian cooperative jamming signals limit the decoding performance of the legitimate receiver as well, and the s.d.o.f.~achieved is still zero. Positive s.d.o.f.~can be obtained by either structured signaling \cite{structured_codes_he_yener_2014_journal} or non-i.i.d.~Gaussian signaling \cite{bassily_ergodic_align}. The exact optimal sum s.d.o.f.~of the wiretap channel with $M$ helpers and the $K$-user multiple access wiretap channel are established to be $\frac{M}{M+1}$ and $\frac{K(K-1)}{K(K-1)+1}$, respectively in \cite{jianwei_ulukus_one_hop}, when full eavesdropper's CSIT is available. In this paper, we show that without eavesdropper's CSIT, the optimal s.d.o.f.~for the wiretap channel with $M$ helpers is still $\frac{M}{M+1}$, while the optimal sum s.d.o.f.~of the $K$-user multiple access wiretap channel decreases to $\frac{K-1}{K}$.

The $K$-user interference channel with an external eavesdropper is studied in \cite{koyluoglu_gamal_lai_poor2011}. When the eavesdropper's CSIT is available, \cite{koyluoglu_gamal_lai_poor2011} proposes a scheme that achieves sum s.d.o.f.~of $\frac{K-1}{2}$. The optimal s.d.o.f.~in this case, however, is established in \cite{jianwei_interference} to be $\frac{K(K-1)}{2K-1}$, using cooperative jamming signals along with interference alignment techniques. When the eavesdropper's CSIT is not available, reference  \cite{koyluoglu_gamal_lai_poor2011} proposes a scheme that achieves a sum s.d.o.f.~of $\frac{K-2}{2}$. In this paper, we establish the optimal s.d.o.f.~in this case to be $\frac{K-1}{2}$.  
A related line of research investigates the wiretap channel, the multiple access wiretap channel, and the broadcast channel with an \emph{arbitrarily varying} eavesdropper \cite{XHe_Yener2010,XHe_Yener2013,broadcast_he_yener}, when the eavesdropper CSIT is not available. The eavesdropper's channel is assumed to be arbitrary, without any assumptions on its distribution, and security is guaranteed for \emph{every} realization of the eavesdropper's channel. This models an exceptionally strong eavesdropper, which may control its own channel in an adversarial manner.  Hence, the optimal sum s.d.o.f.~is zero in each case with single antenna terminals, since the eavesdropper's channel realizations may be exactly equal to the legitimate user's channel realizations. On the other hand, in our model, the eavesdropper's channel gains are drawn from a known distribution, though the realizations are not known at the transmitters. We show that, with this mild assumption, strictly positive s.d.o.f.~can be achieved even with single antennas at each transmitter and receiver for \emph{almost all} channel realizations for helper, multiple access, and interference networks.

\section{System Model and Definitions}

In this paper, we consider three fundamental channel models: the wiretap channel with helpers, the multiple access wiretap channel, and the interference channel with an external eavesdropper. For each channel model, we consider two scenarios of channel variation:
a) fixed channel gains, and b) fading channel gains.
For the case of fixed channel gains, we assume that the channel gains are non-zero and have been drawn independently from a continuous distribution with bounded support and remain fixed for the duration of the communication. On the other hand, in the fading scenario, we assume that the channel gains are non-zero and are drawn from a common continuous distribution with bounded support in an i.i.d.~fashion in each channel use. The common continuous distribution is known at all the terminals in the system.

Let $\Omega$ denote the collection of all channel gains in $n$ channel uses. We assume full CSI at the receivers, that is, both the legitimates receivers and the eavesdropper know $\Omega$. In the following subsections we describe each channel model and provide the relevant definitions.

\subsection{Wiretap Channel with Helpers}

\begin{figure}[t]
\centering
\includegraphics[width=0.6\linewidth]{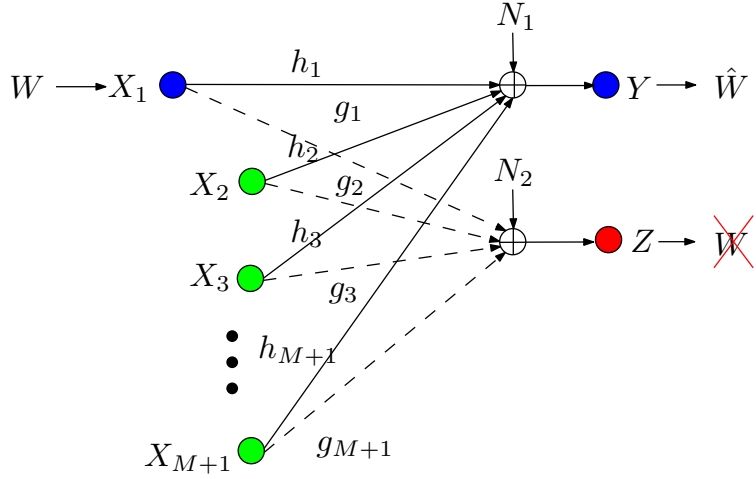}
\caption{Wiretap channel with $M$ helpers.}
\label{fig:wt_model}
\end{figure}

The wiretap channel with $M$ helpers, see Fig.~\ref{fig:wt_model}, is described by,
\begin{align}
Y(t) =& h_1(t)X_1(t) + \sum\limits_{i=2}^{M+1}h_i(t)X_i(t)+ N_1(t)\label{eq:wt_model1}\\
Z(t) =& g_1(t)X_1(t)+\sum\limits_{i=2}^{M+1}g_i(t)X_i(t) +N_2(t)\label{eq:wt_model2}
\end{align}
where $X_1(t)$ denotes the channel input of the legitimate transmitter, and $Y(t)$ denotes the channel output at the legitimate receiver, at time $t$. $X(i), i=2,\ldots, M+1$, are the channel inputs of the $M$ helpers, and $Z(t)$ denotes the channel output at the eavesdropper, at time $t$. In addition, $N_1(t)$ and $N_2(t)$ are white Gaussian noise variables with zero-mean and unit-variance. Here, $h_i(t)$, $g_i(t)$ are the channel gains of the users to the legitimate receiver and the eavesdropper, respectively, and $g_i(t)$s are not known at any of the transmitters. All channel inputs are subject to the average power constraint $E[X_i(t)^2] \leq P$, $i=1,\ldots,M+1$.

The legitimate transmitter wishes to transmit a  message $W$ which is uniformly distributed in $\mathcal{W}$. A secure rate $R$, with $R = \frac{\log|\mathcal{W}|}{n}$ is achievable if there exists a sequence of codes which satisfy the reliability constraints at the legitimate receiver, namely, $\mbox{Pr}[ W\neq \hat{W}] \leq \epsilon_{n}$, and the secrecy constraint, namely,
\begin{align}
\frac{1}{n} I(W;Z^n, \Omega) \leq \epsilon_n
\end{align}
where $\epsilon_n \rightarrow 0$ as $n \rightarrow \infty$. The supremum of all achievable secure rates $R$ is the secrecy capacity $C_s$ and the s.d.o.f., $d_s$, is defined as
\begin{align}
d_s = \lim\limits_{P\rightarrow \infty} \frac{C_s}{\frac{1}{2}\log P}
\end{align}

\subsection{Multiple Access Wiretap Channel}

\begin{figure}[t]
\centering
\includegraphics[width=0.7\linewidth]{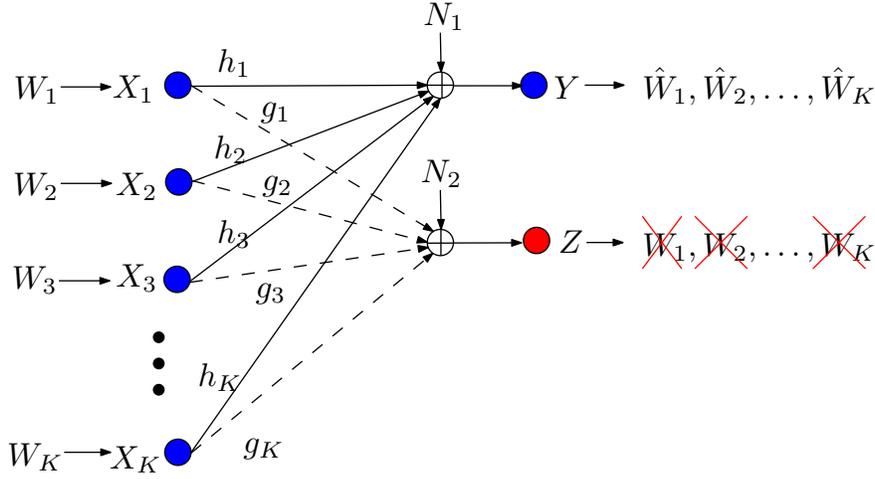}
\caption{$K$-user multiple access wiretap channel.}
\label{fig:mac_model}
\end{figure}

The $K$-user multiple access wiretap channel, see Fig.~\ref{fig:mac_model}, is described by,
\begin{align}
Y(t) =& \sum\limits_{i=1}^{K}h_i(t)X_i(t)+ N_1(t)\label{eq:mac_model1}\\
Z(t) =& \sum\limits_{i=1}^{K}g_i(t)X_i(t) +N_2(t)\label{eq:mac_model2}
\end{align}
where $X_i(t)$ denotes the $i$th user's channel input, $Y(t)$ denotes the legitimate receiver's channel output, and $Z(t)$ denotes the eavesdropper's channel output, at time $t$. In addition, $N_1(t)$ and $N_2(t)$ are white Gaussian noise variables with zero-mean and unit-variance. Here, $h_i(t)$, $g_i(t)$ are the channel gains of the users to the legitimate receiver and the eavesdropper, respectively, and $g_i(t)$s are not known at any of the transmitters. All channel inputs are subject to the average power constraint $E[X_i(t)^2]\leq P$, $i=1,\ldots,K$.

The $i$th user transmits message $W_i$ which is uniformly distributed in $\mathcal{W}_i$. A secure rate tuple $(R_{1},\ldots,R_{K})$, with $R_{i} = \frac{\log|\mathcal{W}_i|}{n}$ is achievable if there exists a sequence of codes which satisfy the reliability constraints at the legitimate receiver, namely, $\mbox{Pr}[ W_{i}\neq \hat{W}_{i}] \leq \epsilon_{n}$, for $i=1,\ldots,K$, and the secrecy constraint, namely,
\begin{align}
\frac{1}{n} I(W_{1}^K;Z^n, \Omega) \leq \epsilon_n
\end{align}
where $\epsilon_n \rightarrow 0$ as $n \rightarrow \infty$. Here, $W_1^K$ denotes the set of all the messages, i.e., $\left\lbrace W_1,\ldots,W_K \right\rbrace$.  An s.d.o.f.~tuple $\left(d_1,\ldots, d_K \right) $ is said to be achievable if a rate tuple $\left(R_1,\ldots, R_K \right) $ is achievable with $d_i = \lim\limits_{P\rightarrow \infty} \frac{R_i}{\frac{1}{2}\log P}$. The sum s.d.o.f., $d_s$, is the largest achievable $\sum_{i=1}^{K} d_i$.

\subsection{Interference Channel with External Eavesdropper}

\begin{figure}[t]
\centering
\includegraphics[width=0.58\linewidth]{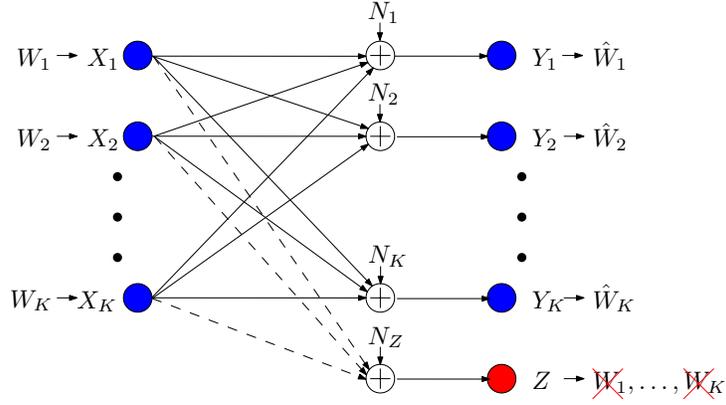}
\caption{$K$-user interference channel with an external eavesdropper.}
\label{fig:interference_model}
\end{figure}

The $K$-user interference channel with an external eavesdropper, see Fig.~\ref{fig:interference_model}, is described by
\begin{align}
Y_i(t) =& \sum_{j=1}^K h_{ji}(t)X_j(t) + N_i(t), \quad i=1,\ldots, K \label{eq:inter_model1}\\
Z(t) =& \sum_{j=1}^K g_{j}(t)X_j(t) + N_Z(t) \label{eq:inter_model2}
\end{align}
where $Y_i(t)$ is the channel output of receiver $i$, $Z(t)$ is the channel output at the eavesdropper, $X_j(t)$ is the channel input of transmitter $j$, $h_{ji}(t)$ is the channel gain from transmitter $j$ to receiver $i$, $g_j(t)$ is the channel gain from transmitter $j$ to the eavesdropper, and $\left\lbrace N_1(t),\ldots, N_K(t), N_Z(t) \right\rbrace$ are mutually independent zero-mean unit-variance white Gaussian noise random variables, at time $t$. The channel gains to the eavesdropper, $g_i(t)$s are not known at any of the transmitters. All channel inputs are subject to the average power constraint $E[X_i(t)^2]\leq P$, $i=1,\ldots,K$.

Transmitter $i$ wishes to send a message $W_i$, chosen uniformly from a set $\mathcal{W}_i$, to receiver $i$. The messages $W_1,\ldots,W_K$ are mutually independent. A secure rate tuple $(R_{1},\ldots,R_{K})$, with $R_{i} = \frac{\log|\mathcal{W}_i|}{n}$ is achievable if there exists a sequence of codes which satisfy the reliability constraints at all the legitimate receivers, namely, $\mbox{Pr}[ W_{i}\neq \hat{W}_{i}] \leq \epsilon_{n}$, for $i=1,\ldots,K$, and the security condition
\begin{align}
\frac{1}{n}I(W_1^K; Z^n, \Omega) \leq \epsilon_n
\end{align}
where $\epsilon_n \rightarrow 0$, as $n\rightarrow \infty$. An s.d.o.f.~tuple $\left(d_1,\ldots, d_K \right) $ is said to be achievable if a rate tuple $\left(R_1,\ldots, R_K \right) $ is achievable with $d_i = \lim\limits_{P\rightarrow \infty} \frac{R_i}{\frac{1}{2}\log P}$. The sum s.d.o.f., $d_s$, is the largest achievable $\sum_{i=1}^{K} d_i$.

\section{Main Results and Discussion}

In this section, we state the main results of this paper. We have the following theorems:
\begin{Theo}\label{theo:wt}
For the wiretap channel with $M$ helpers and no eavesdropper CSIT, the optimal sum s.d.o.f., $d_s$, is given by,
\begin{align}
d_s = \frac{M}{M+1}
\end{align}
almost surely, for both fixed and fading channel gains.
\end{Theo}
\begin{Theo}\label{theo:mac}
For the $K$-user multiple access wiretap channel with no eavesdropper CSIT, the optimal sum s.d.o.f., $d_s$, is given by,
\begin{align}
d_s = \frac{K-1}{K}
\end{align}
almost surely, for both fixed and fading channel gains.
\end{Theo}
\begin{Theo}\label{theo:interference}
For the $K$-user interference channel with an external eavesdropper with no eavesdropper CSIT, the optimal sum s.d.o.f., $d_s$, is given by,
\begin{align}
d_s = \frac{K-1}{2}
\end{align}
almost surely, for both fixed and fading channel gains.
\end{Theo}

We present the proofs of Theorems~\ref{theo:wt} and  \ref{theo:mac} in Section~\ref{sec:theo_wt_mac} and the proof of Theorem~\ref{theo:interference} in Section~\ref{sec:inter_proof}. Let us first state a corollary obtained from Theorems~\ref{theo:wt} and \ref{theo:mac}, which establishes the entire s.d.o.f.~region of the $K$-user multiple access wiretap channel with no eavesdropper CSIT.

\begin{Cor}\label{cor1}
The s.d.o.f.~region of the $K$-user multiple access wiretap channel with no eavesdropper CSIT is given by,
\begin{align}
d_i\geq 0, ~i=1,\ldots,K, \quad \mbox{and} \quad \sum_{i=1}^{K}d_i \leq& \frac{K-1}{K} \label{eq:region}
\end{align}
\end{Cor}

The proof of Corollary~\ref{cor1} follows directly from Theorems~\ref{theo:wt} and \ref{theo:mac}. In particular, we can treat the $K$-user multiple access wiretap channel as a $(K-1)$ helper wiretap channel with transmitter $i$ as the legitimate transmitter, and the remaining transmitters as helpers. This achieves the corner points $d_i = \frac{K-1}{K}$ and $d_j=0$ for $j \neq i$ from Theorem~\ref{theo:wt}. Therefore, given the sum s.d.o.f.~upper bound in Theorem~\ref{theo:mac}, and that each corner point with s.d.o.f.~of $\frac{K-1}{K}$ for a single user is achievable, the region in Corollary~\ref{cor1} follows.

It is useful, at this point, to compare our results to the cases when the eavesdropper's CSI is available at the transmitter. Table~\ref{table:comparison} shows a comparison of the optimal s.d.o.f.~values with and without eavesdropper CSIT. Interestingly, there is no loss in s.d.o.f.~for the wiretap channel with helpers due to the absence of eavesdropper's CSIT.

\begin{table}[t]
\centering
\renewcommand{\arraystretch}{1.5}
\begin{tabular}{ |>{\centering}m{6cm}|>{\centering}m{4cm}|>{\centering}m{4cm}| }
\hline
\textbf{Channel model}~& \textbf{With Eve CSIT} & \textbf{Without Eve CSIT}
\tabularnewline \hline
Wiretap channel with $M$ helpers  &   $\frac{M}{M+1}$  & $\frac{M}{M+1}$\tabularnewline \hline
$K$-user multiple access wiretap channel & $\frac{K(K-1)}{K(K-1)+1}$  & $\frac{K-1}{K}$ \tabularnewline \hline
$K$-user interference channel with an external eavesdropper & $\frac{K(K-1)}{2K-1}$ & $\frac{K-1}{2}$\tabularnewline \hline
\end{tabular}
\caption{Summary of s.d.o.f.~values with and without eavesdropper CSIT.}
\label{table:comparison}
\end{table}

However, for the multiple access wiretap channel and the interference channel with an external eavesdropper, the optimal s.d.o.f.~decreases due to the unavailability of eavesdropper CSIT. For the multiple access wiretap channel, as the number of users, $K$ increases, the optimal sum s.d.o.f.~approaches $1$ as $\sim \frac{1}{K^2}$ with eavesdropper's CSIT but only as $\sim \frac{1}{K}$ without eavesdropper's CSIT. Therefore, the loss of s.d.o.f.~as a fraction of the optimal sum s.d.o.f.~with eavesdropper CSIT is $\sim \frac{1}{K}$ for large $K$.

For the interference channel with an external eavesdropper too, there is a loss in s.d.o.f.~due to the unavailability of the eavesdropper's CSIT. However, in this case, the optimal s.d.o.f.~without eavesdropper CSIT closely tracks the s.d.o.f.~with eavesdropper CSIT. In fact, it can be verified that the s.d.o.f.~loss is bounded by $\frac{1}{4}$, which implies that the loss of s.d.o.f.~as a fraction of the optimal s.d.o.f.~with eavesdropper CSIT is $\sim \frac{1}{K}$ for large $K$, in this case also.

For the multiple access wiretap channel, we also consider the case where some of the transmitters have the eavesdropper's CSI. We state our achievable s.d.o.f.~in this case in the following theorem.

\begin{Theo}\label{theo:m_mac}
In the $K$-user MAC-WT, where $1\leq m\leq K$ transmitters have eavesdropper CSI, and the remaining $K-m$ transmitters have no eavesdropper CSI, the following sum s.d.o.f.~is achievable,
\begin{align}
d_s = \frac{m(K-1)}{m(K-1)+1}
\end{align}
almost surely, for both fixed and fading channel gains.
\end{Theo}

We present the proof of Theorem~\ref{theo:m_mac} in Section~\ref{sec:theo_mac2_proof}. In this case, we note that when only one user has eavesdropper CSIT, i.e., $m=1$, our achievable rate is the same as when no user has eavesdropper CSIT as in  Theorem~\ref{theo:mac}. On the other hand, when all users have eavesdropper CSIT, i.e., $m=K$, our achievable rate is the same as the optimal sum s.d.o.f.~in \cite{jianwei_ulukus_one_hop}. We note that our achievable sum s.d.o.f.~varies from the no eavesdropper CSIT result in Theorem~\ref{theo:mac} to the full eavesdropper CSIT sum s.d.o.f.~in \cite{jianwei_ulukus_one_hop} as $m$ increases from  $1$ to $K$.

\section{Proofs of Theorems~\ref{theo:wt} and  \ref{theo:mac}}\label{sec:theo_wt_mac}

First, we note that an achievable scheme for Theorem~\ref{theo:wt} implies an achievable scheme for Theorem~\ref{theo:mac}, since the $K$-user multiple access wiretap channel may be treated as a wiretap channel with $(K-1)$ helpers. Further, we note that a converse for Theorem~\ref{theo:mac} suffices as a converse for Theorem~\ref{theo:wt}. Thus, we will only provide achievable schemes for Theorem~\ref{theo:wt} and a converse proof for Theorem~\ref{theo:mac}. An alternate converse for  Theorem~\ref{theo:wt} also follows from the converse presented in \cite{jianwei_ulukus_one_hop} for the wiretap channel with $M$ helpers and with eavesdropper CSIT, as the converse for the case of known eavesdropper CSIT serves as a converse for the case of unknown eavesdropper CSIT.

\subsection{Achievability for the Wiretap Channel with Helpers}

We now present achievable schemes for the wiretap channel with $M$ helpers for both fixed and fading channels. We begin with the case of fixed channel gains.

\subsubsection{Fixed Channel Gains}

For fixed channels, we use the technique of real interference alignment \cite{real_inter_align,real_inter_align_exploit}. Let $\{V_2,V_3,\cdots,\\V_{M+1},U_1,U_2,U_3,\cdots,U_{M+1}\}$ be mutually independent discrete random variables, each of which  uniformly drawn from the same PAM constellation $C(a,Q)$
\begin{equation}
C(a,Q) = a \{ -Q, -Q+1, \ldots, Q-1,Q\} \label{eq:constellation}
\end{equation}
where $Q$ is a positive integer and $a$ is a real number used to normalize the transmission power, and is also the minimum distance between the points belonging to $C(a,Q)$. Exact values of $a$ and $Q$ will be specified later. We choose the input signal of the legitimate transmitter as
\begin{equation}
X_1  = \frac{1}{h_1}U_1 +  \sum_{k=2}^{M+1} \alpha_k  V_k
\label{legit-signal}
\end{equation}
where $\{\alpha_k\}^{M+1}_{k=2}$ are rationally independent among themselves and also rationally independent of all channel gains. The input signal of the $j$th helper, $j=2,\cdots,M+1$, is chosen as
\begin{equation}
X_j = \frac{1}{h_j} U_j
\label{helper-signal}
\end{equation}
Note that, neither the legitimate transmitter signal in (\ref{legit-signal}) nor the helper signals in (\ref{helper-signal}) depend on the eavesdropper CSI $\{g_k\}_{k=1}^{M+1}$. With these selections, observations of the receivers are given by,
\begin{align}
Y & = \sum_{k=2}^{M+1} {h_1 \alpha_k} V_k + \left( \sum_{j=1}^{M+1} U_j \right)+ N_1 \label{eq:real_align1}\\
Z & = \sum_{k=2}^{M+1} {g_1 \alpha_k} V_k +  \sum_{j=1}^{M+1} \frac{g_j}{h_j} U_j + N_2 \label{eq:real_align2}
\end{align}

\begin{figure*}[t]
\centering
\includegraphics[scale=0.8]{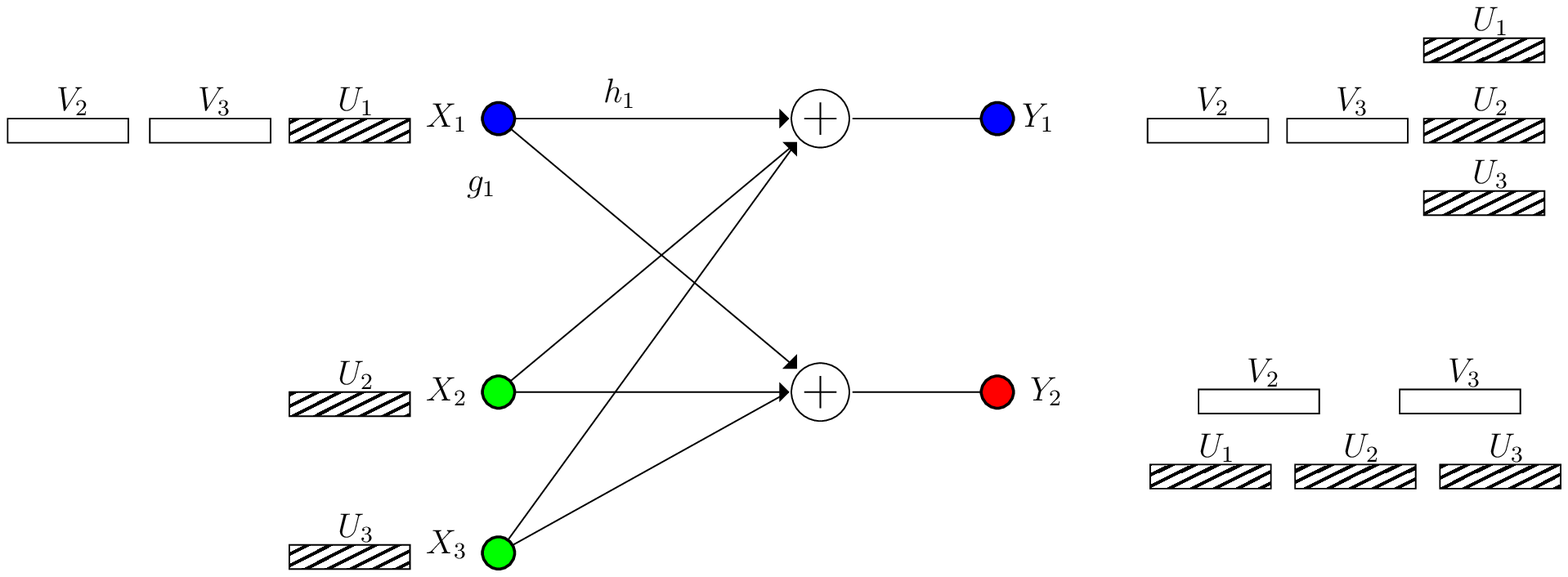}
\caption{Illustration of the alignment scheme for the Gaussian wiretap channel with $M$ helpers with no eavesdropper CSI.}
\label{fig:gwc_no_csi_one_helper_no_csi_ia}
\end{figure*}

The intuition here is as follows: We use $M$ independent sub-signals $V_k$, $k=2,\cdots,M+1$, to represent the original message $W$. The input signal $X_1$ is a linear combination of $V_k$s and a jamming signal $U_1$. At the legitimate receiver, all of the cooperative jamming signals, $U_k$s, are
aligned such that they occupy a small portion of the signal space. Since $\left\{1, h_1 \alpha_2, h_1 \alpha_3, \cdots, h_1 \alpha_{M+1}\right\}$ are rationally independent for  all channel gains, except for a set of Lebesgue measure zero, the signals $\left\{V_2,V_3,\cdots,V_{M+1}, \sum_{j=1}^{M+1} U_j \right\}$ can be distinguished by the legitimate receiver. In addition, we observe that $\left\{\frac{g_1}{h_1}, \cdots, \frac{g_{M+1}}{h_{M+1}}\right\}$ are rationally independent, and therefore, $\left\{U_1,U_2,\cdots,U_{M+1}\right\}$ \emph{span} the \emph{entire space} at the eavesdropper; see Fig.~\ref{fig:gwc_no_csi_one_helper_no_csi_ia}.  Here, by the \emph{entire space}, we mean the maximum number of \emph{dimensions} that the eavesdropper is capable of decoding, which is $(M+1)$ in this case. Since the entire space at the eavesdropper is occupied by the cooperative jamming signals, the message signals $\{V_2, V_3, \cdots, V_{M+1}\}$ are secure, as we will mathematically prove in the sequel.

The following secrecy rate is achievable \cite{csiszar}
\begin{equation}
C_s \ge I(\mathbf{V};Y) - I(\mathbf{V};Z)
\label{eqn:gwcnc-lower-bound}
\end{equation}
where $\mathbf{V} \stackrel{\Delta}{=} \{V_2,V_3,\cdots, V_{M+1}\}$. Note that since $\Omega$ is known at both the legitimate receiver and the eavesdropper, it can be considered to be an additional output at both the legitimate receiver and the eavesdropper. Further, since $\mathbf{V}$ is chosen to be independent of $\Omega$, $\Omega$ should appear in the conditioning of each of the mutual information quantities in \eqref{eqn:gwcnc-lower-bound}. We keep this in mind, but drop it for the sake of notational simplicity.

First, we  use Fano's inequality to bound the first term in \eqref{eqn:gwcnc-lower-bound}. Note that the space observed at receiver $1$ consists of $(2Q+1)^M (2MQ+2Q+1)$ points in $(M+1)$ dimensions, and the sub-signal in each dimension is drawn from a constellation of $C(a,(M+1)Q)$. Here, we use the property that $C(a,Q)\subset C(a,(M+1)Q)$.  By using the Khintchine-Groshev theorem of Diophantine approximation in number theory \cite{real_inter_align, real_inter_align_exploit}, we can bound the minimum distance $d_{min}$ between the points
in receiver 1's space as follows: For any $\delta>0$, there exists a constant $k_\delta$ such that
\begin{equation}
\label{eqn:gwch_lb_of_d_m_helper}
d_{min} \ge \frac{ k_\delta  a}{((M+1)Q)^{M+\delta}}
\end{equation}
for almost all rationally independent $\left\{1, h_1 \alpha_2, h_1 \alpha_3, \cdots, h_1 \alpha_{M+1}\right\}$, except for a set of Lebesgue measure zero. Then, we can upper bound the
probability of decoding error of such a PAM scheme by considering the additive
Gaussian noise at receiver $1$,
\begin{align}
\mathbb{P}\left[\mathbf{V}\neq\hat{\mathbf{V}} \right]
& \le \exp\left(  - \frac{d_{min}^2}{8}\right)   \\
& \le \exp\left(  -
\frac{a^2k_\delta^2}{8 ((M+1)Q)^{2(M+\delta)}}\right)
\end{align}
where $\hat{\mathbf{V}}$ is the estimate of $\mathbf{V}$ by choosing the closest point in the constellation based on observation $Y$. For any $\delta>0$, if we choose $Q = P^{\frac{1-\delta}{2(M+1+\delta)}}$ and $a=\gamma P^{\frac{1}{2}}/Q$, where  $\gamma$
is a constant independent of $P$, then
\begin{align}
\mathbb{P}\left[\mathbf{V}\neq\hat{\mathbf{V}}\right]
& \le \exp\left( -\frac{k_\delta^2 \gamma^2 (M+1)^2 P}{8 ((M+1)Q)^{2(M+\delta)+2}} \right)
\\ &
 = \exp\left( -\frac{k_\delta^2 \gamma^2 (M+1)^2 P^\delta}{8 (M+1)^{2(M+1+\delta)}} \right)
\end{align}
and we can have $\mathbb{P}\left[\mathbf{V}\neq\hat{\mathbf{V}}\right] \to 0$  as $P\to\infty$. To
satisfy the power constraint at the transmitters, we can simply choose
\begin{align}
& \gamma  \le \min\left\{\left[\frac{1}{|h_1| } + \sum_{k=2}^{M+1} |\alpha_k|\right]^{-1},|h_2|, |h_3|, \cdots,
|h_{M+1}|\right\}
\end{align}
By Fano's inequality and the Markov chain $\mathbf{V}\rightarrow
Y\rightarrow\hat{\mathbf{V}}$, we know that
\begin{align}
 H(\mathbf{V} | Y)
&\le H(\mathbf{V}|\hat{\mathbf{V}}) \\
&\le 1 +
\exp\left( -\frac{k_\delta^2 \gamma^2 (M+1)^2 P^\delta}{8 (M+1)^{2(M+1+\delta)}}
\right) \log(2Q+1)^M
\\
& = o(\log P)
\label{eqn:nocsi_fano_logp}
\end{align}
where $\delta$ and $\gamma$  are fixed, and  $o(\cdot)$ is the little-$o$ function. This means that
\begin{align}
  I(\mathbf{V};Y)
& = H(\mathbf{V}) - H(\mathbf{V}|Y) \\
& \ge
 H(\mathbf{V}) - o(\log P) \\
& =
 \log(2Q+1)^M - o(\log P) \\
& \ge  \log P^{\frac{M(1-\delta)}{2(M+1+\delta)}} - o(\log P) \\
& =  {\frac{M(1-\delta)}{M+1+\delta}} \left( \frac{1}{2} \log P \right) - o(\log P)
 \label{eqn:gwch_wiretap_m_helper_lb_ixy1}
\end{align}

Next, we need to bound the second term in \eqref{eqn:gwcnc-lower-bound},
\begin{align}
I(\mathbf{V};Z)
& = I(\mathbf{V},\mathbf{U};Z) - I(\mathbf{U};Z|\mathbf{V}) \\
& = I(\mathbf{V},\mathbf{U};Z) - H(\mathbf{U}|\mathbf{V}) +  H(\mathbf{U}|{Z},\mathbf{V}) \\
& = I(\mathbf{V},\mathbf{U};Z) - H(\mathbf{U}) +  H(\mathbf{U}|{Z},\mathbf{V})
\\
& = h(Z) - h(Z|\mathbf{V},\mathbf{U})
 - H(\mathbf{U}) +  H(\mathbf{U}|{Z},\mathbf{V})
\\
& = h(Z) - h(N_2)- H(\mathbf{U}) +  H(\mathbf{U}|{Z},\mathbf{V})
\\
& \le h(Z) - h(N_2)- H(\mathbf{U}) +  o(\log P)
\label{eqn:gwch_nocsi_decode_U_given_Y_and_V}
\\
& \le \frac{1}{2} \log P - \frac{1}{2} \log 2 \pi e
 - \log(2Q+1)^{M+1} +  o(\log P) \label{eq:upper_bound_hz}\\
& \le \frac{1}{2} \log  P  - \frac{(M+1)(1-\delta)}{2(M+1+\delta)}\log P +
o(\log P) \\
& =  \frac{(M+2)\delta}{M+1+\delta}\left( \frac{1}{2}\log P \right) +
o(\log P)
\label{eqn:gwch_wiretap_m_helper_lb_ixy2}
\end{align}
where $\mathbf{U}\stackrel{\Delta}{=}\{U_1,U_2,\cdots,U_{M+1}\}$, and \eqref{eqn:gwch_nocsi_decode_U_given_Y_and_V} is due to the  fact that given $\mathbf{V}$ and $Z$, the eavesdropper can decode $\mathbf{U}$ with probability of error
approaching zero since $\left\{\frac{g_1}{h_1}, \cdots, \frac{g_{M+1}}{h_{M+1}}\right\}$ are rationally independent for all channel gains, except for a set of Lebesgue measure zero. Then, by Fano's inequality, $H(\mathbf{U}|Z,\mathbf{V})\le o(\log P)$ similar to the step in \eqref{eqn:nocsi_fano_logp}. In addition,  $h(Z)\leq \frac{1}{2}\log P + o(\log P)$ in \eqref{eq:upper_bound_hz}, since all the channel gains are drawn from a known distribution with bounded support.

Combining \eqref{eqn:gwch_wiretap_m_helper_lb_ixy1} and \eqref{eqn:gwch_wiretap_m_helper_lb_ixy2}, we have
\begin{align}
C_s
& \ge I(\mathbf{V};Y) - I(\mathbf{V};Z) \\
& \ge  {\frac{M(1-\delta)}{M+1+\delta}} \left( \frac{1}{2} \log P \right)
 - \frac{(M+2)\delta}{M+1+\delta}\left( \frac{1}{2}\log P \right) -
o(\log P)
 \\
& ={\frac{M - (2M+2)\delta}{M+1+\delta}} \left(\frac{1}{2}\log P\right) - o(\log P)\label{eq:real_align_final}
\end{align}
where again $o(\cdot)$ is the little-$o$ function. If we choose $\delta$ arbitrarily small, then we can achieve $\frac{M}{M+1}$ s.d.o.f.~for this model where there is no eavesdropper CSI at the transmitters.

\subsubsection{Fading Channel Gains}

Now, we present an achievable scheme for the case of fading channel gains, i.e., when the channel gains vary in an i.i.d.~fashion from one time slot to another.
In this scheme, the legitimate transmitter sends $M$ independent Gaussian symbols, $\mathbf{V} = \left\{ V_2,\ldots,V_{M+1} \right\}$ securely to the legitimate receiver in $(M+1)$ time slots. This is done as follows:

At time $t=1,\ldots,M+1$, the legitimate transmitter sends a scaled artificial noise, i.e., cooperative jamming, symbol $U_1$ along with information symbols as,
\begin{align}
X_1(t) = \frac{1}{h_1(t)}U_1 + \sum_{k=2}^{M+1} \alpha_k(t) V_k
\end{align}
where the $\alpha_{k}(t)$s are chosen such that the $(M+1)\times (M+1)$ matrix $T$, with entries $T_{ij}= \alpha_i(j)h_1(j)$, where $\alpha_1(j) = \frac{1}{h_1(j)}$, is full rank. The $j$th helper, $j=2,\ldots,M+1$, transmits:
\begin{align}
X_j(t) = \frac{1}{h_j(t)}U_j
\end{align}
The channel outputs at time $t$ are,
\begin{align}
Y(t) & = \sum_{k=2}^{M+1} {h_1(t) \alpha_k(t)} V_k + \left( \sum_{j=1}^{M+1} U_j \right) +N_1(t)\label{eq:vector_align1} \\
Z(t) & = \sum_{k=2}^{M+1} {g_1(t) \alpha_k(t)} V_k +  \sum_{j=1}^{M+1} \frac{g_j(t)}{h_j(t)} U_j + N_2(t)\label{eq:vector_align2}
\end{align}

Note the similarity of the scheme with that of the real interference scheme for fixed channel gains, i.e., the similarity between \eqref{eq:vector_align1}-\eqref{eq:vector_align2} and \eqref{eq:real_align1}-\eqref{eq:real_align2}. Indeed the alignment structure after $(M+1)$ channel uses is exactly as in Fig.~\ref{fig:gwc_no_csi_one_helper_no_csi_ia}. Note also how the artificial noise symbols align at the legitimate receiver over $(M+1)$ time slots. At high SNR, at the end of the $(M+1)$ slots, the legitimate receiver recovers $(M+1)$ linearly independent equations with $(M+1)$ variables: $V_2,\ldots, V_{M+1}, \sum_{j=1}^{M+1}U_j$. Thus, the legitimate receiver can recover $\mathbf{V} \stackrel{\Delta}{=} \left( V_2,\ldots,V_{M+1} \right)$ within noise variance.

Formally, let us define $\mathbf{U} \stackrel{\Delta}{=} \left(U_1,\ldots,U_{M+1}\right)$, $\mathbf{Y} \stackrel{\Delta}{=} \left(Y(1),\ldots,Y(M+1)\right)$, and $\mathbf{Z} \stackrel{\Delta}{=} \left(Z(1),\ldots,Z(M+1)\right)$. The observations at the legitimate receiver and the eavesdropper can then be compactly written as
\begin{align}
\mathbf{Y} = \left(\mathbf{A}_V,\mathbf{A}_U\right)\left(\begin{array}{c}\mathbf{V}^T\\\mathbf{U}^T
\end{array}\right) + \mathbf{N}_1\\
\mathbf{Z} = \left(\mathbf{B}_V,\mathbf{B}_U\right)\left(\begin{array}{c}\mathbf{V}^T\\\mathbf{U}^T
\end{array}\right) + \mathbf{N}_2
\end{align}
where $\mathbf{A}_V$ is a $(M+1)\times M$ matrix with $\left(\mathbf{A}_V\right)_{ij} = h_1(i)\alpha_{j+1}(i)$,
$\mathbf{A}_{U}$ is a $(M+1)\times (M+1) $ matrix with all ones, $\mathbf{B}_V$ is a $(M+1)\times M$ matrix with $\left(\mathbf{B}_V\right)_{ij} = g_1(i)\alpha_{j+1}(i)$, and $\mathbf{B}_U$ is a $(M+1)\times (M+1)$ matrix with $\left(\mathbf{B}_V\right)_{ij} =\frac{g_{j}(i)}{h_{j}(i)}$. $\mathbf{N}_1$ and $\mathbf{N}_2$ are $(M+1)$ dimensional vectors containing the noise variables $N_1(t)$ and $N_2(t)$, respectively, for $t=1,\ldots,M+1$. To calculate differential entropies, we use the following lemma.
\begin{Lem}\label{lem:dif_entropy}
Let $\mathbf{A}$ be an $M\times N$ dimensional matrix and let $\mathbf{X} = \left(X_1,\ldots,X_N\right)^T$ be a jointly Gaussian random vector with zero-mean and variance $P\mathbf{I}$. Also, let $\mathbf{N} = \left(N_1,\ldots, N_M\right)^T$ be a jointly Gaussian random vector with zero-mean and variance $\sigma^2\mathbf{I}$, independent of $\mathbf{X}$. If $r=\mbox{rank}(\mathbf{A})$, then,
\begin{align}
h(\mathbf{A}\mathbf{X} + \mathbf{N}) = r\left(\frac{1}{2}\log P\right) + o(\log P)
\end{align}
\end{Lem}
We present the proof of Lemma~\ref{lem:dif_entropy} in Appendix~\ref{a:dif_entropy}.

Using Lemma~\ref{lem:dif_entropy}, we compute
\begin{align}
I(\mathbf{V};\mathbf{Y}) =& h(\mathbf{Y}) - h(\mathbf{Y}|\mathbf{V})\\
=& (M+1)\frac{1}{2}\log P - h(\mathbf{A}_U \mathbf{U}^T + \mathbf{N}_1)+o(\log P)\label{eq:step1}\\
=& (M+1)\left(\frac{1}{2}\log P\right)  - \frac{1}{2}\log P + o(\log P)\label{eq:step2}\\
=& M \left(\frac{1}{2}\log P\right) +o(\log P)
\end{align}
where \eqref{eq:step1} follows since $\mathbf{U}$ and $\mathbf{N}_1$ are independent of $\mathbf{V}$ and since $\left(\mathbf{A}_V,\mathbf{A}_U\right)$ has rank $(M+1)$, and \eqref{eq:step2} follows since $\mathbf{A}_U$ has rank $1$.
We also have,
\begin{align}
I(\mathbf{V};\mathbf{Z}) =& h(\mathbf{Z}) - h(\mathbf{Z}|\mathbf{V})\\
=& (M+1)\frac{1}{2}\log P - h(\mathbf{B}_U\mathbf{U}^T + \mathbf{N}_2)+o(\log P)\\
=& (M+1)\frac{1}{2}\log P - (M+1)\frac{1}{2}\log P + o(\log P)\\
=& o(\log P)\label{eq:secrecy}
\end{align}
where we have used the fact that both $(\mathbf{B}_V,\mathbf{B}_U)$ and $\mathbf{B}_U$ have rank $(M+1)$, almost surely. Note that, in both calculations above, we have implicitly used the fact that $\Omega$ is known to both the legitimate receiver and the eavesdropper, and that it appears in the conditioning of each mutual information and differential entropy term.
Equation \eqref{eq:secrecy} means that the leakage to the eavesdropper does not scale with $\log P$.

Now, consider the vector wiretap channel from $\mathbf{V}$ to $\mathbf{Y}$ and $\mathbf{Z}$, by treating the $K$ slots in the scheme above as one channel use. Similar to \eqref{eqn:gwcnc-lower-bound}, the following secrecy rate is achievable
\begin{align}
 C_s^{vec} \geq& I(\mathbf{V};\mathbf{Y}) - I(\mathbf{V};\mathbf{Z})\\
 =& M\left(\frac{1}{2}\log P\right) + o(\log P)
\end{align}
Since each channel use of this vector channel uses $(M+1)$ actual channel uses, the achievable rate for the actual channel is,
\begin{align}
C_s \geq \frac{M}{M+1} \left(\frac{1}{2} \log P\right) + o(\log P)\label{eq:vec_align_final}
\end{align}
Thus, the achievable s.d.o.f.~of this scheme is $\frac{M}{M+1}$. The results in \eqref{eq:real_align_final} and \eqref{eq:vec_align_final} complete the achievability of Theorem~\ref{theo:wt}, for fixed and fading channel gains, respectively.

\subsection{Converse for the Multiple Access Wiretap Channel}\label{converse:mac}

We combine techniques from \cite{jianwei_ulukus_one_hop} and \cite{aligned_image_sets_jafar} to prove the converse. Here, we use $\mathbf{X}_i$ to denote the collection of all channel inputs $\{ X_i(t), ~t=1,\ldots,n\}$ of transmitter $i$. Similarly, we use $\mathbf{Y}$ and $\mathbf{Z}$ to denote the channel outputs at the legitimate receiver and the eavesdropper, respectively, over $n$ channel uses. We further define $\mathbf{X}_1^K$ as the collection of all channel inputs from all of the transmitters, i.e., $\{\mathbf{X}_i, ~i=1\ldots, K\}$. Finally, for a fixed $j$, we use $\mathbf{X}_{-j}$ to denote all channel inputs from all transmitters except transmitter $j$, i.e., $\{\mathbf{X}_i, ~i\neq j, ~i=1\ldots,K\}$. Since all receivers know $\Omega$, it appears in the conditioning in every entropy and mutual information term below. We keep this in mind, but drop it for the sake of notational simplicity. We divide the proof into three steps.

\subsubsection{Deterministic Channel Model}\label{sec:det_model}

We will show that there is no loss of s.d.o.f.~in considering the following integer-input integer-output deterministic channel in \eqref{eq:detmod1}-\eqref{eq:detmod2} instead of the one in \eqref{eq:mac_model1}-\eqref{eq:mac_model2}
\begin{align}
Y(t) =& \sum\limits_{i=1}^{K} \lf h_{i}(t)X_i(t) \rf\label{eq:detmod1}\\
Z(t) =& \sum\limits_{i=1}^{K}\lf g_i(t)X_i(t) \rf \label{eq:detmod2}
\end{align}
with the constraint that
\begin{align}
X_{i} \in \left\{ 0,1,\ldots, \lf \sqrt{P} \rf \right\}\label{eq:pow_constraints}
\end{align}
To that end, we will show that given any codeword tuple $(\mathbf{X}^G_1,\ldots,\mathbf{X}^G_K)$ for the original channel of \eqref{eq:mac_model1}-\eqref{eq:mac_model2}, we can construct a codeword tuple $(\mathbf{X}^D_1,\ldots,\mathbf{X}^D_K)$ with $X^D_i(t) = \lf X^G_i(t) \rf \mbox{ mod } \lfloor \sqrt{P} \rfloor$, for the deterministic channel of \eqref{eq:detmod1}-\eqref{eq:detmod2}, that achieves an s.d.o.f.~no smaller than the s.d.o.f.~achieved by $(\mathbf{X}^G_1,\ldots,\mathbf{X}^G_K)$ on the original channel. Let us denote by $\mathbf{Y}^G$ and $\mathbf{Z}^G$, the outputs of the  original channel of \eqref{eq:mac_model1}-\eqref{eq:mac_model2}, when  $(\mathbf{X}^G_1,\ldots,\mathbf{X}^G_K)$ is the input, that is,
\begin{align}
Y^G(t) \stackrel{\Delta}{=} \sum\limits_{i=1}^K h_i(t)X^G_i(t) + N_1(t)\\
Z^G(t) \stackrel{\Delta}{=} \sum\limits_{i=1}^K g_i(t)X^G_i(t) + N_2(t)
\end{align}
Similarly, define
\begin{align}
Y^D(t) \stackrel{\Delta}{=}& \sum\limits_{i=1}^{K} \lf h_{i}(t)X^D_i(t) \rf\\
Z^D(t) \stackrel{\Delta}{=}& \sum\limits_{i=1}^{K}\lf g_i(t)X^D_i(t) \rf
\end{align}
It suffices to show that
\begin{align}
I(W_i;\mathbf{Y}^G) \leq& I(W_i;\mathbf{Y}^D) +no(\log P)\label{eq:dof}\\
I(W_1^K;\mathbf{Z}^D) \leq& I(W_1^K;\mathbf{Z}^G) +no(\log P)\label{eq:sec}
\end{align}
for every $i=1,\ldots,K$. Here, \eqref{eq:dof} states that the information rate to the legitimate receiver in the discretized channel is at least as large as the information rate in the original Gaussian channel, and \eqref{eq:sec} states that the information leakage to the eavesdropper in the discretized channel is at most at the level of the information leakage in the original Gaussian channel, both of which quantified within a $o(\log P)$.

The proof of \eqref{eq:dof} follows along similar lines as the proof presented in \cite{aligned_image_sets_jafar} and is omitted here. To prove \eqref{eq:sec}, we first define
\begin{align}
\bar{Z}(t) \stackrel{\Delta}{=}& \sum_{i=1}^K \lf g_i(t)\lf X^G_i(t)\rf\rf \\
\hat{Z}(t) \stackrel{\Delta}{=}& \bar{Z}(t) - Z^D(t)\\
\tilde{Z}(t) \stackrel{\Delta}{=}& \lf Z^G(t)\rf - \bar{Z}(t) -\lf N_2(t)\rf
\end{align}
Then, we have,
\begin{align}
I(W_1^K;\mathbf{Z}^D) \leq& I(W_1^K;\mathbf{Z}^D,\mathbf{Z}^G,\bar{\mathbf{Z}})\\
=& I(W_1^K;\mathbf{Z}^G) + I(W_1^K; \bar{\mathbf{Z}}|\mathbf{Z}^G) + I(W_1^K;\mathbf{Z}^D|\bar{\mathbf{Z}},\mathbf{Z}^G)\\
\leq& I(W_1^K;\mathbf{Z}^G) + H(\bar{\mathbf{Z}}|\mathbf{Z}^G) + H(\mathbf{Z}^D|\bar{\mathbf{Z}},\mathbf{Z}^G)  \\
\leq& I(W_1^K;\mathbf{Z}^G) + H(\bar{\mathbf{Z}}|\lfloor \mathbf{Z}^G \rfloor) + H(\mathbf{Z}^D|\bar{\mathbf{Z}})\\
\leq& I(W_1^K;\mathbf{Z}^G) + H(\bar{\mathbf{Z}}|\bar{\mathbf{Z}}  + \tilde{\mathbf{Z}}+\lf \mathbf{N}_2 \rf) + H(\hat{\mathbf{Z}})\\
\leq& I(W_1^K;\mathbf{Z}^G) + \sum_{i=1}^n H({\bar{Z}(t)}|\bar{Z}(t)  + \tilde{{Z}}(t)+\lf N_2(t)\rf)+ \sum_{i=1}^n H(\hat{{Z}}(t))\label{eq:last_but final}\\
\leq& I(W_1^K;\mathbf{Z}^G) +no(\log P) \label{eq:final}
\end{align}
where $\lfloor \mathbf{Z}^G\rfloor = \left(\lfloor Z^G(1)\rfloor,\ldots,\lfloor Z^G(n)\rfloor\right)$. Here, \eqref{eq:final} follows since $H(\hat{{Z}}(t)) \leq o(\log P)$ following the steps of the proof in \cite[Appendix~A.2]{aligned_image_sets_jafar}. In addition, recalling that $\Omega$ appears in the conditioning of each term in \eqref{eq:last_but final}, note that $H({\bar{Z}(t)}|\bar{Z}(t)  + \tilde{{Z}}(t)+\lf N_2(t)\rf, \Omega) \leq E\left[H({\bar{Z}(t)}|\bar{Z}(t)  + \tilde{{Z}}(t)+\lf N_2(t)\rf,g_1^K =\tilde{g}_1^K)\right] \leq o(\log P)$ using \cite[Lemma~E.1, Appendix~E]{avestimehr_diggavi_tse}, since $\tilde{Z}(t)$ is integer valued and is bounded by $\sum_{i=1}^K \tilde{g}_i(t) + K+1$ for each realization $\tilde{g}_i(t)$ of $g_i(t)$.

Therefore, the s.d.o.f.~of the deterministic channel in \eqref{eq:detmod1}-\eqref{eq:detmod2} with integer channel inputs as described in \eqref{eq:pow_constraints} is no smaller than the s.d.o.f.~of the original channel in \eqref{eq:mac_model1}-\eqref{eq:mac_model2}. Consequently, any upper bound (e.g., converse) developed for the s.d.o.f.~of \eqref{eq:detmod1}-\eqref{eq:detmod2} will serve as an upper bound for the s.d.o.f.~of \eqref{eq:mac_model1}-\eqref{eq:mac_model2}.  Thus, we will consider this deterministic channel in the remaining part of the converse.

\subsubsection{An Upper Bound on the Sum Rate}

We begin as in the \emph{secrecy penalty lemma} in \cite{jianwei_ulukus_one_hop}, i.e., \cite[Lemma~1]{jianwei_ulukus_one_hop}. Note that, unlike \cite[Lemma~1]{jianwei_ulukus_one_hop}, channel inputs are integer here and satisfy \eqref{eq:pow_constraints}:
\begin{align}
n \sum_{i=1}^{K}R_i \leq& I(W_1^K;\mathbf{Y}) - I(W_1^K;\mathbf{Z}) +n\epsilon\\
\leq& I(W_1^K;\mathbf{Y}|\mathbf{Z}) +n\epsilon\\
\leq& I(\mathbf{X}_1^K;\mathbf{Y}|\mathbf{Z}) +n\epsilon\\
\leq& H(\mathbf{Y}|\mathbf{Z}) +n\epsilon\label{eq:dep1}\\
=& H(\mathbf{Y},\mathbf{Z}) - H(\mathbf{Z}) +n\epsilon\\
\leq& H(\mathbf{X}_1^K, \mathbf{Y},\mathbf{Z})- H(\mathbf{Z}) +n\epsilon\\
=& H(\mathbf{X}_1^K)- H(\mathbf{Z}) +n\epsilon\label{eq:dep}\\
\leq& \sum\limits_{k=1}^{K} H(\mathbf{X}_k) -H(\mathbf{Z})  +n\epsilon\label{eq:inter}
\end{align}
where \eqref{eq:dep} follows since $H(\mathbf{Y},\mathbf{Z}|\mathbf{X}_1^K)=0$ for the channel in \eqref{eq:detmod1}-\eqref{eq:detmod2}.
Also, to ensure decodability at the legitimate receiver, we use the \emph{role of a helper lemma} in \cite{jianwei_ulukus_one_hop}, i.e., \cite[Lemma~2]{jianwei_ulukus_one_hop},
\begin{align}
n \sum\limits_{i\neq j}R_i \leq& I(W_{-j};\mathbf{Y}) +n\epsilon'\\
\leq& I(\mathbf{X}_{-j};\mathbf{Y})+n\epsilon'\\
=& H(\mathbf{Y}) - H(\mathbf{Y}|\mathbf{X}_{-j})+n\epsilon'\\
=& H(\mathbf{Y}) - H(\lf \mathbf{h}_j\mathbf{X}_j\rf)+n\epsilon' \label{eq:decode_step1}\\
=& H(\mathbf{Y}) - H(\lf \mathbf{h}_j\mathbf{X}_j\rf, \mathbf{X}_j)+ H(\mathbf{X}_j|\lf \mathbf{h}_j\mathbf{X}_j\rf)+n\epsilon'\\
\leq& H(\mathbf{Y}) - H(\mathbf{X}_j)+ H(\mathbf{X}_j|\lf \mathbf{h}_j\mathbf{X}_j\rf)+n\epsilon'\\
\leq& H(\mathbf{Y}) - H(\mathbf{X}_j)+ \sum_{t=1}^{n} H(X_j(t)|\lf {h}_j(t)X_j(t)\rf)+n\epsilon'\label{eq:decode_step2}\\
\leq& H(\mathbf{Y}) - H(\mathbf{X}_j) + n\epsilon'+nc \label{eq:decode1}
\end{align}
where $\mathbf{h}_j\mathbf{X}_j \stackrel{\Delta}{=} \left\lbrace h_j(t)X_j(t), t=1,\ldots,n \right\rbrace $, and recalling that $\Omega$ appears in the conditioning of each term in \eqref{eq:decode_step2}, \eqref{eq:decode1} follows using the following lemma.
\begin{Lem}\label{lem:inter_step}
Let $X$ be an integer valued random variable satisfying \eqref{eq:pow_constraints}, and $h$ be drawn from a distribution $F(h)$ satisfying $\int_{-\infty}^{\infty} \log \left(1+\frac{1}{|h|}\right) dF(h) \leq c$ for some $c \in \mathbb{R}$. Then,
\begin{align}
H(X|\lf hX \rf,h) \leq c
\end{align}
\end{Lem}
The proof of this lemma is presented in Appendix~\ref{a:proof_of_lemma_inter_step}. The constraint imposed in Lemma~\ref{lem:inter_step} is a mild technical condition. It can be verified that a sufficient condition for satisfying the constraint is that there exists an $\epsilon >0$ such that the probability density function (pdf) is bounded in the interval $(-\epsilon,\epsilon)$. Most common distributions such as Gaussian, exponential and Laplace satisfy this condition.

Eliminating $H(\mathbf{X}_j)$s using \eqref{eq:inter} and \eqref{eq:decode1}, we get,
\begin{align}
Kn\sum\limits_{i=1}^{K} R_i \leq& KH(\mathbf{Y}) - H(\mathbf{Z})+nK(\epsilon'+c)+n\epsilon\\
\leq& (K-1)\frac{n}{2}\log P + \left(H(\mathbf{Y}) - H(\mathbf{Z})\right)+n\epsilon'' \qquad \hspace*{-0.9cm}
\end{align}
where $\epsilon'' =  o(\log P)$. Dividing by $n$ and letting $n\rightarrow \infty$,
\begin{align}
K \sum\limits_{i=1}^{K} R_i\leq& (K-1)\frac{1}{2}\log P +  \epsilon''+\lim_{n\rightarrow \infty} \frac{1}{n}\left(H(\mathbf{Y}) - H(\mathbf{Z})\right)
\end{align}
Now dividing by $\frac{1}{2}\log P$ and taking $P\rightarrow \infty$,
\begin{align}
\sum\limits_{i=1}^{K} d_i \leq& \frac{K-1}{K} + \frac{1}{K}\lim_{P \rightarrow \infty}\lim_{n\rightarrow \infty} \frac{H(\mathbf{Y}) - H(\mathbf{Z})}{\frac{n}{2}\log P}\label{eq:rate_bound}
\end{align}

\subsubsection{Bounding the Difference of Entropies}\label{sec:mac_entropy_bound}

We now upper bound the difference of entropies $H(\mathbf{Y}) - H(\mathbf{Z})$ in (\ref{eq:rate_bound}) as:
\begin{align}
H(\mathbf{Y}) - H(\mathbf{Z}) \leq& \sup_{\{\mathbf{X}_i\}:\mathbf{X}_i \indep \mathbf{X}_j} H(\mathbf{Y}) - H(\mathbf{Z}) \label{eq:max1}\\
\leq& \sup_{\{\mathbf{X}_i\}} ~ H(\mathbf{Y}) - H(\mathbf{Z}) \label{eq:max2}
\end{align}
where $X \indep Y$ is used to denote that $X$ and $Y$ are statistically independent and \eqref{eq:max2} follows from \eqref{eq:max1} by relaxing the condition of independence in \eqref{eq:max1}. Since the $\mathbf{X}_i$s in \eqref{eq:max2} may be arbitrarily correlated, we can think of the $K$ single antenna terminals as a single transmitter with $K$ antennas. Thus, we wish to maximize $H(\mathbf{Y}) - H(\mathbf{Z})$, where $\mathbf{Y}$ and $\mathbf{Z}$ are two single antenna receiver outputs, under the constraint that the channel gains to $\mathbf{Z}$ are unknown at the transmitter. This brings us to the $K$-user MISO broadcast channel setting of \cite{aligned_image_sets_jafar}. We know from \cite[eqns.~(75)-(103)]{aligned_image_sets_jafar} that even without any security or decodability constraints, the difference of entropies, $H(\mathbf{Y}) - H(\mathbf{Z})$ cannot be larger than $no(\log P)$, if the channel gains to the second receiver are unknown. Thus,
\begin{align}
H(\mathbf{Y}) - H(\mathbf{Z}) \leq no(\log P) \label{eq:entropy_bound}
\end{align}

Using \eqref{eq:entropy_bound} in \eqref{eq:rate_bound}, we have
\begin{align}
\sum\limits_{i=1}^{K} d_i \leq& \frac{K-1}{K}
\end{align}
This completes the converse proof of Theorem~\ref{theo:mac}.

\section{Proof of Theorem~\ref{theo:interference}}\label{sec:inter_proof}

In this section, we present the proof of Theorem~\ref{theo:interference}. We first present separate achievable schemes for fixed and fading channel gains and then present the converse. For the interference channel, we require asymptotic schemes with both real \cite{real_inter_align_exploit}, and vector space alignment \cite{cadambe_jafar_interference} techniques. The converse combines techniques from \cite{jianwei_ulukus_interference_2013} and \cite{aligned_image_sets_jafar}.

\subsection{Achievability for the Interference Channel}

An achievable scheme for the interference channel with an external eavesdropper and no eavesdropper CSIT is presented in \cite[Theorem~3]{koyluoglu_gamal_lai_poor2011}. That scheme achieves sum s.d.o.f.~of  $\frac{K-2}{2}$. Here, we present the optimal schemes which achieve $\frac{K-1}{2}$ sum s.d.o.f. In this section, we focus on the case when $K=3$, which highlights the main ideas of the general $K$-user scheme and present the $K$-user scheme in Appendix~\ref{k_user_schemes}. As in the achievability for the wiretap channel with helpers, we use the techniques of real and vector space alignment for fixed channel gains and fading channel gains, respectively. However, unlike the case of wiretap channel with helpers, we need to use asymptotic alignment in each case. We begin with the case of fixed channel gains.

\subsubsection{Fixed Channel Gains}

\begin{figure}[t]
\centering
\includegraphics[width=\linewidth]{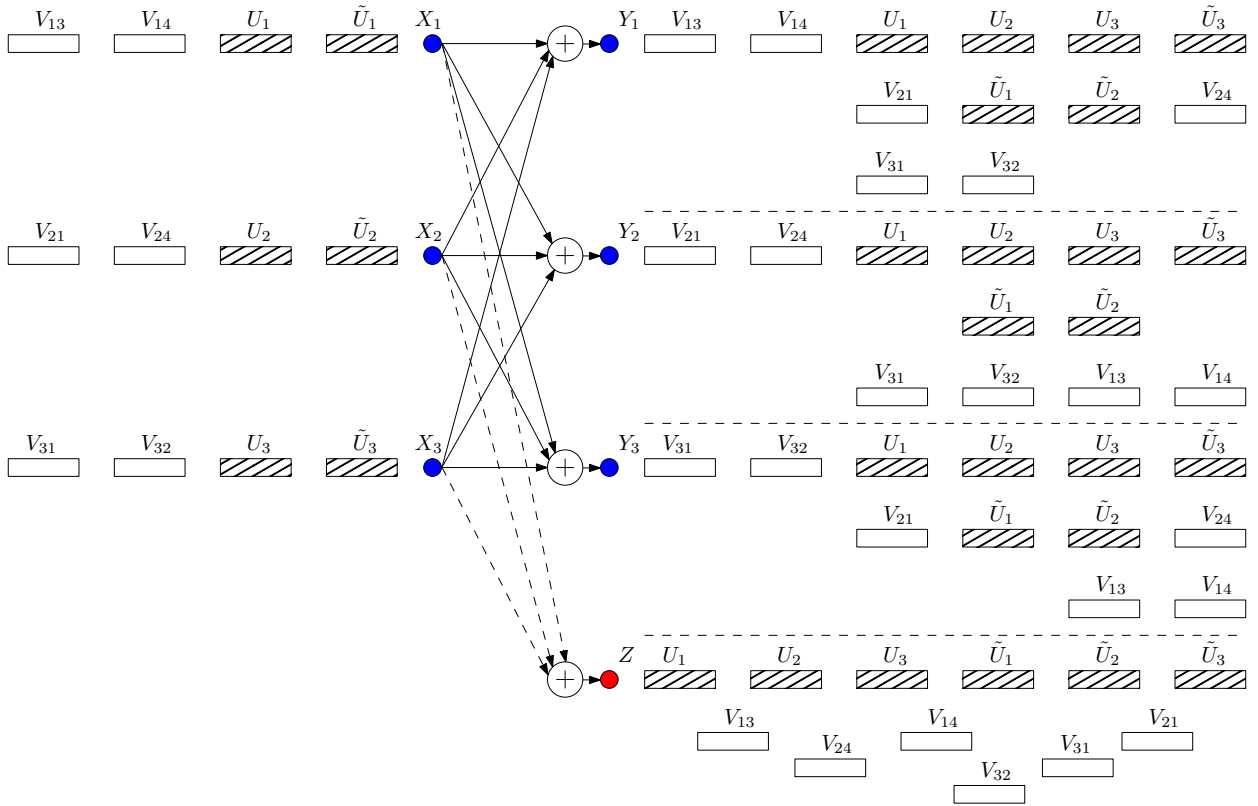}
\caption{Alignment for the interference channel with $K=3$.}
\label{fig:alignment_inter}
\end{figure}

We use the technique of asymptotic real interference alignment introduced in \cite{real_inter_align_exploit}. Fig.~\ref{fig:alignment_inter} shows the desired signal alignment at the receivers and the eavesdropper. In the figure, the boxes labeled by $V$ denote the message symbols, while the hatched boxes labeled with $U$ denote artificial noise symbols. We observe from Fig.~\ref{fig:alignment_inter} that $4$ out of $6$ \emph{signal dimensions} are buried in the artificial noise. Thus, heuristically, the s.d.o.f.~for each legitimate user pair is $\frac{2}{6}=\frac{1}{3}$, and the sum s.d.o.f.~is, therefore, $3\times \frac{1}{3} =1$, as expected from our optimal sum s.d.o.f.~expression $\frac{K-1}{2} = \frac{3-1}{2}=1$.

In the $K$-user case, we have a similar alignment scheme. Each transmitter sends two artificial noise blocks along with $(K-1)$ message blocks. At each legitimate receiver, the $2K$ noise blocks from the $K$ transmitters align such that they occupy only $(K+1)$ \emph{block dimensions}. This is done by aligning $\tilde{U}_{k}$ with $U_{k+1}$ for $k=1,\ldots, (K-1)$, at each legitimate receiver. The unintended messages at each legitimate receiver are aligned underneath the $(K+1)$ artificial noise dimensions. To do so, we use two main ideas. First, two blocks from the same transmitter cannot be aligned at \emph{any} receiver. This is because if two blocks from the same transmitter align at any receiver, they align at every other receiver as well, which is clearly not desirable. Secondly, each message block aligns with the same artificial noise block at every unintended receiver. Thus, in Fig.~\ref{fig:alignment_inter}, $V_{21}$ and $V_{24}$ appear in different columns at each receiver. Further, $V_{21}$ appears underneath $U_1$ at both of the unintended legitimate receivers $1$ and $2$. It can be verified that these properties hold for every message block. As an interesting by-product, this alignment scheme provides confidentiality of the unintended messages at the legitimate transmitters for free. The $(K-1)$ intended message blocks at a legitimate receiver occupy distinct block dimensions; thus, achieving a d.o.f.~of $\frac{K-1}{2K}$ for each transmitter-receiver pair.  At the eavesdropper, no alignment is possible since its CSIT is unavailable. Thus, the $2K$ artificial noise blocks occupy the full space of $2K$ block dimensions. This ensures security of the messages at the eavesdropper.

Note that we require two artificial noise blocks to be transmitted from each transmitter. When the eavesdropper CSIT is available, the optimal achievable scheme, presented in \cite{jianwei_interference}, requires one artificial noise block from each transmitter; the $K$ noise blocks from the $K$ transmitters are aligned with the messages at the eavesdropper in order to ensure security. In our case, however, the eavesdropper's CSIT is not available. Thus, in order to guarantee security, we need a total of $2K$ noise blocks to occupy the full space of $2K$ block dimensions at the eavesdropper. This is achieved by sending two artificial noise blocks from each transmitter. Further, to achieve an s.d.o.f.~of $\frac{K-1}{2K}$ per user pair, we need to create $(K-1)$ noise-free message block dimensions at each legitimate receiver. We ensure this by systematically aligning the $2K$ noise symbols to occupy only $(K+1)$ block dimensions at each legitimate receiver. To the best of our knowledge, this is the first achievable scheme in the literature that uses two artificial noise blocks from each transmitter and then aligns them to maximize the noise-free message dimensions at each legitimate receiver.

Let us now present the $3$-user scheme in more detail. Let $m$ be a large integer. Also, let $c_1$, $c_2$, $c_3$ and $c_4$ be real constants drawn from a fixed continuous distribution with bounded support independently of each other and of all the channel gains. This ensures that the $c_i$s are \emph{rationally independent} of each other and of the channel gains. Now, we define four sets $T_i$, $i=1,\ldots,4$, as follows:
\begin{align}
T_1 \stackrel{\Delta}{=}& \left\lbrace h_{11}^{r_{11}}h_{12}^{r_{12}}h_{13}^{r_{13}}h_{21}^{r_{21}}h_{31}^{r_{31}}h_{32}^{r_{32}}h_{23}^{r_{32}}c_1^s:~ r_{jk},s \in \left\lbrace 1,\ldots,m \right\rbrace  \right\rbrace \\
T_2 \stackrel{\Delta}{=}&   \left\lbrace h_{21}^{r_{21}}h_{22}^{r_{22}}h_{23}^{r_{23}}\left(\frac{h_{12}}{h_{11}} \right)^{r_{12}}\left(\frac{h_{13}}{h_{11}} \right)^{r_{13}}h_{31}^{r_{31}}h_{32}^{r_{32}}c_2^s:~r_{jk},s \in \left\lbrace 1,\ldots,m \right\rbrace    \right\rbrace \\
T_3\stackrel{\Delta}{=}& \left\lbrace h_{31}^{r_{31}}h_{32}^{r_{32}}h_{33}^{r_{33}}\left(\frac{h_{21}}{h_{22}} \right)^{r_{21}}\left(\frac{h_{23}}{h_{22}} \right)^{r_{23}}h_{12}^{r_{12}}h_{13}^{r_{13}}c_3^s:~r_{jk},s \in \left\lbrace 1,\ldots,m \right\rbrace  \right\rbrace \\
T_4 \stackrel{\Delta}{=}& \left\lbrace h_{31}^{r_{31}}h_{32}^{r_{32}}h_{33}^{r_{33}}h_{21}^{r_{21}}h_{12}^{r_{12}}h_{13}^{r_{13}}h_{23}^{r_{23}}c_4^s:~ r_{jk},s \in \left\lbrace 1,\ldots,m \right\rbrace  \right\rbrace
\end{align}
Let  $M_i$ be the cardinality of the set $T_i$. Note that all the $M_i$s are the same, which we denote by $M$, which is given as,
\begin{align}
M \stackrel{\Delta}{=} m^8
\end{align}
We subdivide each message $W_i$ into 2 independent sub-messages $V_{ij}, j=1,\ldots,4, j\neq i,i+1$. For each transmitter $i$, let $\mathbf{p}_{ij}$ be the vector containing all the elements of $T_j$, for $j\neq i,i+1$. For any given $(i,j)$ with $j\neq i,i+1$, $\mathbf{p}_{ij}$ represents the dimension along which message $V_{ij}$ is sent. Further, at each transmitter $i$, let $\mathbf{q}_i$ and $\tilde{\mathbf{q}}_{i}$ be vectors containing all the elements in sets $T_{i}$ and $\beta_iT_{i+1}$, respectively, where
\begin{align}
\beta_i = \begin{cases}
\frac{1}{h_{ii}},\quad &\mbox{if } i= 1,2\\
1, &\mbox{if } i=3
\end{cases}
\end{align}
The vectors $\mathbf{q}_i$ and $\tilde{\mathbf{q}}_{i}$ represent dimensions along which artificial noise symbols $U_i$ and $\tilde{U}_i$, respectively, are sent. We define a $4M$ dimensional vector $\mathbf{b}_i$ by stacking the $\mathbf{p}_{ij}$s, $\mathbf{q}_i$ and $\tilde{\mathbf{q}}_i$ as
\begin{align}
\mathbf{b}_i^T = \left[ \mathbf{p}_{i1}^T \ldots \mathbf{p}_{i(i-1)}^T \quad \mathbf{p}_{i(i+2)}^T \ldots \mathbf{p}_{i4}\quad \mathbf{q}_i \quad \tilde{\mathbf{q}}_i \right]
\end{align}
The transmitter encodes $V_{ij}$ using an $M$ dimensional vector $\mathbf{v}_{ij}$, and the cooperative jamming signals $U_i$ and $\tilde{U}_i$ using $M$ dimensional vectors $\mathbf{u}_i$ and $\tilde{\mathbf{u}}_i$, respectively. Each element of $\mathbf{v}_{ij}$, $\mathbf{u}_i$ and $\tilde{\mathbf{u}}_i$ are drawn in an i.i.d.~fashion from $C(a,Q)$ in \eqref{eq:constellation}. Let
\begin{align}
\mathbf{a}_i^T =\left[ \mathbf{v}_{i1}^T \ldots \mathbf{v}_{i(i-1)}^T \quad \mathbf{v}_{i(i+2)}^T \ldots \mathbf{v}_{i4}\quad \mathbf{u}_i \quad \tilde{\mathbf{u}}_i \right]
\end{align}
The channel input of transmitter $i$ is then given by
\begin{align}
x_i = \mathbf{a}_i^T\mathbf{b}
\end{align}

Let us now analyze the structure of the received signals at the receivers. For example, consider receiver $1$. The desired signals at receiver $1$, $\mathbf{v}_{13}$ and $\mathbf{v}_{14}$ arrive along dimensions $h_{11}T_3$ and $h_{11}T_4$, respectively. Since only $T_i$ (and not $T_j, j\neq i$) contains $c_i$, these dimensions are rationally independent. Thus, they appear along different columns in Fig.~\ref{fig:alignment_inter}. The artificial noise symbols $\mathbf{u}_1$, $\mathbf{u}_2$, $\mathbf{u}_3$ and $\tilde{\mathbf{u}}_3$ arrive along dimensions $h_{11}T_1$, $h_{21}T_{2}$, $h_{31}T_3$ and $h_{31}T_4$, respectively. Again they are all rationally separate and thus, appear along different columns in Fig.~\ref{fig:alignment_inter}. Further, they are all separate from the dimensions of the desired signals, because $T_3$ and $T_4$ do not contain $h_{11}$, while $T_1$ and $T_2$ do not contain either $c_3$ or $c_4$. On the other hand, the unintended signals $\mathbf{v}_{21}$ and $\mathbf{v}_{31}$ arrive along $h_{21}T_1$ and $h_{31}T_1$, and since $T_1$ contains powers of $h_{21}$ and $h_{31}$, they align with the artificial noise $\mathbf{u}_1$ in $\tilde{T}_1$, where,
\begin{align}
\tilde{T}_1 \stackrel{\Delta}{=}& \left\lbrace h_{11}^{r_{11}}h_{12}^{r_{12}}h_{13}^{r_{13}}h_{21}^{r_{21}}h_{31}^{r_{31}}h_{32}^{r_{32}}h_{23}^{r_{32}}c_1^s:~ r_{jk},s \in \left\lbrace 1,\ldots,m+1 \right\rbrace  \right\rbrace
\end{align}
Similarly, we define
\begin{align}
\tilde{T}_2 \stackrel{\Delta}{=}&   \left\lbrace h_{21}^{r_{21}}h_{22}^{r_{22}}h_{23}^{r_{23}}\left(\frac{h_{12}}{h_{11}} \right)^{r_{12}}\left(\frac{h_{13}}{h_{11}} \right)^{r_{13}}h_{31}^{r_{31}}h_{32}^{r_{32}}c_2^s:~r_{jk},s \in \left\lbrace 1,\ldots,m+1 \right\rbrace    \right\rbrace \\
\tilde{T}_3\stackrel{\Delta}{=}& \left\lbrace h_{31}^{r_{31}}h_{32}^{r_{32}}h_{33}^{r_{33}}\left(\frac{h_{21}}{h_{22}} \right)^{r_{21}}\left(\frac{h_{23}}{h_{22}} \right)^{r_{23}}h_{12}^{r_{12}}h_{13}^{r_{13}}c_3^s:~r_{jk},s \in \left\lbrace 1,\ldots,m+1 \right\rbrace  \right\rbrace \\
\tilde{T}_4 \stackrel{\Delta}{=}& \left\lbrace h_{31}^{r_{31}}h_{32}^{r_{32}}h_{33}^{r_{33}}h_{21}^{r_{21}}h_{12}^{r_{12}}h_{13}^{r_{13}}h_{23}^{r_{23}}c_4^s:~ r_{jk},s \in \left\lbrace 1,\ldots,m+1 \right\rbrace  \right\rbrace
\end{align}
We note that the unintended signals $\mathbf{v}_{32}$ and $\mathbf{v}_{24}$ arrive along $h_{31}T_2$ and $h_{21}T_4$ and thus, align with $\mathbf{u}_{2}$ and $\tilde{\mathbf{u}}_{3}$, respectively, in $\tilde{T}_2$ and $\tilde{T}_4$. Thus, they appear in the same column in Fig.\ref{fig:alignment_inter}. Finally, the artificial noise symbols $\tilde{\mathbf{u}}_1$ and $\tilde{\mathbf{u}}_2$ align with $\mathbf{u}_2$ and $\mathbf{u}_3$, respectively.

At receiver $2$, the desired signals $\mathbf{v}_{21}$ and $\mathbf{v}_{24}$ arrive along rationally independent dimensions $h_{22}T_1$ and $h_{22}T_4$, respectively. The artificial noise symbols $\mathbf{u}_1$, $\mathbf{u}_2$, $\mathbf{u}_3$ and $\tilde{\mathbf{u}}_3$ arrive along dimensions $h_{12}T_1$, $h_{22}T_{2}$, $h_{32}T_3$ and $h_{32}T_4$, respectively. Thus, they lie in dimensions $\tilde{T}_1$, $\tilde{T_2}$, $\tilde{T}_{3}$ and $\tilde{T}_4$, respectively. They are all separate from the dimensions of the desired signals, because $\tilde{T}_1$ and $\tilde{T}_4$ do not contain $h_{22}$, while $\tilde{T}_2$ and $\tilde{T}_3$ do not contain either $c_1$ or $c_4$. The artificial noise symbols $\tilde{\mathbf{u}}_1$ and $\tilde{\mathbf{u}}_2$ arrive along dimensions $\left(\frac{h_{12}}{h_{11}}\right)T_2$ and $T_3$, respectively; thus, they align with $\mathbf{u}_{2}$ and $\mathbf{u}_3$ in $\tilde{T}_2$ and $\tilde{T}_3$, respectively. The unintended signals $\mathbf{v}_{13}$ and $\mathbf{v}_{14}$ arrive along $h_{12}T_3$ and $h_{12}T_4$, respectively, and lie in $\tilde{T}_3$ and $\tilde{T}_4$, respectively. Similarly, $\mathbf{v}_{31}$ and $\mathbf{v}_{32}$ lie in $\tilde{T}_1$ and $\tilde{T}_2$, respectively. A similar analysis is true for receiver $3$ as well.

At the eavesdropper, there is no alignment, since the channel gains of the eavesdropper are not known at the transmitters. In fact, the artificial noise symbols all arrive along different dimensions at the receiver. Thus, heuristically, they exhaust the decoding capability of the eavesdropper almost completely.

We note that the interference at each receiver is confined to the dimensions $\tilde{T}_1$, $\tilde{T}_2$, $\tilde{T}_3$ and $\tilde{T}_4$. Further, these dimensions are separate from the dimensions occupied by the desired signals at each receiver. Specifically, at receiver $i$, the desired signals occupy dimensions $h_{ii}T_{j}, j\neq i,i+1$. These dimensions are separate from $\tilde{T}_i$ and $\tilde{T}_{i+1}$, since only $T_j$ contains powers of $c_j$. Further, $\tilde{T}_j, j\neq i,i+1$ do not contain powers of $h_{ii}$. Thus, the set
\begin{align}
S = \left( \bigcup_{j\neq i,i+1}h_{ii}T_j \right)\bigcup \left(\bigcup_{j=1}^{4}\tilde{T}_{j} \right)
\end{align}
has cardinality
\begin{align}
M_S = 2m^{8} + 4(m+1)^8
\end{align}
Intuitively, out of these $M_S$ dimensions, $2m^8$ dimensions carry the desired signals. Thus, the s.d.o.f.~of each legitimate user pair is $\frac{2m^8}{2m^8+4(m+1)^8}$ which approaches $\frac{1}{3}$ as $m\rightarrow\infty$. Thus, the sum s.d.o.f.~is $1$. We omit the formal calculation of the achievable rate here and instead present it in Appendix~\ref{a:fixed} for the general $K$-user case.    Further, note that the unintended messages at each receiver are buried in artificial noise, see Fig.~\ref{fig:alignment_inter}. Thus, our scheme provides confidentiality of messages from unintended legitimate receivers as well.

\subsubsection{Fading Channel Gains}

Our scheme uses asymptotic vector space alignment introduced in \cite{cadambe_jafar_interference}.
Let $\Gamma = (K-1)^2=(3-1)^2 = 4$. We use $M_n = 2n^\Gamma + 4(n+1)^\Gamma$ channel uses to transmit $6n^\Gamma$ message symbols securely to the legitimate receivers in the presence of the eavesdropper. Thus, we achieve a sum s.d.o.f.~of $\frac{6n^\Gamma}{2n^\Gamma + 4(n+1)^\Gamma}$, which approaches $1$ as $n \rightarrow \infty$.

First, at transmitter $i$, we divide its message $W_i$ into $2$ sub-messages $V_{ij}, j=1,\ldots,4, j\neq i,i+1$. Each $V_{ij}$ is encoded into $n^{\Gamma}$ independent streams $v_{ij}(1),\ldots,v_{ij}(n^\Gamma)$, which we denote as $\mathbf{v}_{ij} \stackrel{\Delta}{=} \left( v_{ij}(1),\ldots,v_{ij}(n^\Gamma)\right)^T $. We also require artificial noise symbols $U_i$ and $\tilde{U}_i$ at each transmitter $i$. We encode the artificial noise symbols $U_i$ and $\tilde{U}_i$ as
\begin{align}
\mathbf{u}_i \stackrel{\Delta}{=}& \left( u_i(1),\ldots, u_i((n+1)^\Gamma)\right)^T, i=1,2,3\\
\tilde{\mathbf{u}}_i \stackrel{\Delta}{=}& \left(\tilde{u}_i(1),\ldots,\tilde{u}_i(n^\Gamma)\right)^T, i=1,2\\
\tilde{\mathbf{u}}_3 \stackrel{\Delta}{=}& \left(\tilde{u}_i(1),\ldots,\tilde{u}_i((n+1)^\Gamma)\right)^T
\end{align}
In each channel use $t\leq M_n$, we choose precoding column vectors $\mathbf{p}_{ij}(t)$, $\mathbf{q}_i(t)$ and $\tilde{\mathbf{q}}_i(t)$  with the same number of elements as $\mathbf{v}_{ij}$, $\mathbf{u}_i$ and $\tilde{\mathbf{u}}_i$, respectively. In channel use $t$, transmitter $i$ sends
\begin{align}
X_i(t) = \sum\limits_{j\neq i,i+1} \mathbf{p}_{ij}(t)^T\mathbf{v}_{ij} + \mathbf{q}_i(t)^T\mathbf{u}_i + \tilde{\mathbf{q}}_i(t)^T\tilde{\mathbf{u}}_i
\end{align}
where we have dropped the limits on $j$ in the summation for notational simplicity.
By stacking the precoding vectors for all $M_n$ channel uses, we let,
\begin{align}
\mathbf{P}_{ij} = \left(\begin{array}{c}
\mathbf{p}_{ij}(1)^T\\
\vdots\\
\mathbf{p}_{ij}^T(M_n)
\end{array} \right), \qquad \mathbf{Q}_{i} = \left(\begin{array}{c}
\mathbf{q}_{i}(1)^T\\
\vdots\\
\mathbf{q}_{i}(M_n)^T
\end{array} \right), \qquad \tilde{\mathbf{Q}}_{i} = \left(\begin{array}{c}
\tilde{\mathbf{q}}_{i}(1)^T\\
\vdots\\
\tilde{\mathbf{q}}_{i}(M_n)^T
\end{array} \right)
\end{align}
Now, letting $\mathbf{X}_i = \left(X_i(1),\ldots,X_i(M_n) \right)^T $, the channel input for transmitter $i$ over $M_n$ channel uses can be compactly represented as
\begin{align}
\mathbf{X}_i = \sum\limits_j \mathbf{P}_{ij}\mathbf{v}_{ij} + \mathbf{Q}_i\mathbf{u}_i + \tilde{\mathbf{Q}}_i\tilde{\mathbf{u}}_i
\end{align}

Recall that, channel use $t$, the channel output at receiver $l$ and the eavesdropper are, respectively, given by
\begin{align}
Y_l(t) =& \sum\limits_{k=1}^{3} h_{kl}(t)X_k(t) + N_l(t) \\
Z(t) =& \sum\limits_{k=1}^{3} g_{k}(t)X_k(t) +N_{Z}(t)
\end{align}
where we have dropped the Gaussian noise at high SNR. Let $\mathbf{H}_{kl} \stackrel{\Delta}{=} \mbox{diag}\left(h_{kl}(1),\ldots,h_{kl}(M_n) \right)$. Similarly, define $\mathbf{G}_{k} = \mbox{diag}\left(g_{k}(1),\ldots,g_{k}(M_n)\right) $. The channel outputs at receiver $l$ and the eavesdropper over all $M_n$ channel uses, $\mathbf{Y}_l = \left(Y_l(1),\ldots,Y_l(M_n) \right)^T$ and $\mathbf{Z} = \left(Z(1),\ldots,Z(M_n) \right)^T$, respectively, can be represented by
\begin{align}
\mathbf{Y}_l =& \sum\limits_{k=1}^{3} \mathbf{H}_{kl}\mathbf{X}_{k}+\mathbf{N}_l\\
=& \sum\limits_{k=1}^3 \mathbf{H}_{kl}\left( \sum\limits_{\substack{j=1\\j\neq k,k+1}}^{4} \mathbf{P}_{kj}\mathbf{v}_{kj} + \mathbf{Q}_k\mathbf{u}_k + \tilde{\mathbf{Q}}_k\tilde{\mathbf{u}}_k \right)+\mathbf{N}_l\\
=&\sum\limits_{\substack{j=1\\j\neq l,l+1}}^{4} \mathbf{H}_{ll}\mathbf{P}_{lj}\mathbf{v}_{lj} + \sum\limits_{\substack{k=1\\k\neq l}}^{3} \sum\limits_{\substack{j=1\\j\neq k,k+1}}^{4} \mathbf{H}_{kl}\mathbf{P}_{kj}\mathbf{v}_{kj} + \sum\limits_{k=1}^{3} \mathbf{H}_{kl}\left( \mathbf{Q}_k\mathbf{u}_k + \tilde{\mathbf{Q}}_k\tilde{\mathbf{u}}_k\right)+\mathbf{N}_l\label{eq:legit_sig_3}\\
\intertext{and,}
 \mathbf{Z} =& \sum\limits_{k=1}^{3} \mathbf{G}_{k}\mathbf{X}_{k}+\mathbf{N}_{Z}\\
=& \sum\limits_{k=1}^{3}\sum\limits_{\substack{j=1\\j\neq k,k+1}}^{4} \mathbf{G}_k\mathbf{P}_{kj}\mathbf{v}_{kj} + \sum\limits_{k=1}^{3} \mathbf{G}_{k}\left( \mathbf{Q}_k\mathbf{u}_k + \tilde{\mathbf{Q}}_k\tilde{\mathbf{u}}_k\right)+\mathbf{N}_{Z}\label{eq:eve_sig_3}
\end{align}

Now, receiver $l$ wants to decode $\mathbf{v}_{lj}, j=1,\ldots,4, j\neq l,l+1$. Thus, the remaining terms in \eqref{eq:legit_sig_3} constitute interference at the $l$th receiver. Let $CS(\mathbf{X})$ denote the column space of matrix $\mathbf{X}$.  Then, $I_l$ denoting the space spanned by this interference is given by
\begin{align}
I_l = \left( \bigcup\limits_{k\neq l,j\neq k,k+1} CS\left( \mathbf{H}_{kl}\mathbf{P}_{kj}\right)\right)   \bigcup \left(\bigcup\limits_{k=1}^3 CS\left( \mathbf{H}_{kl}\mathbf{Q}_k\right)   \right) \bigcup \left(\bigcup\limits_{k=1}^3 CS\left( \mathbf{H}_{kl}\tilde{\mathbf{Q}}_k\right)   \right)
\end{align}
Note that there are $2n^\Gamma$ symbols to be decoded by each legitimate receiver in $2n^\Gamma + 4(n+1)^\Gamma$ channel uses. Thus, for decodability, the interference can occupy a subspace of rank at most $4(n+1)^\Gamma$, that is,
\begin{align}
\mbox{rank}(I_l) \leq 4(n+1)^\Gamma
\end{align}
To that end, we align the noise and message subspaces at each legitimate receiver appropriately. Note that no such alignment is possible at the external eavesdropper since the transmitters do not have its CSI. In addition, note that we have a total of $2n^\Gamma + 4(n+1)^\Gamma$ artificial noise symbols which will span the full received signal space at the eavesdropper and secure all the messages.

Fig.~\ref{fig:alignment_inter} shows the alignment we desire. We remark that the same figure represents the alignment of signals both for real interference alignment and the vector space alignment schemes. Now, let us enumerate the conditions for the desired signal alignment at each receiver. From Fig.~\ref{fig:alignment_inter}, it is clear that there are $6$ alignment equations at each legitimate receiver, corresponding to four unintended messages and two artificial noise symbols $\tilde{U}_1$ and $\tilde{U}_2$. Table~\ref{table:align_eqns} shows the alignment equations for each legitimate receiver.

\begin{table}[t]
\renewcommand{\arraystretch}{1.5}
\centering
\begin{tabular}{|c|c|c|c|c|}
\hline
 & $\mathbf{Q}_1$ & $\mathbf{Q}_2$ & $\mathbf{Q}_3$ & $\tilde{\mathbf{Q}}_3$\\
\hline
\multirow{2}{*}{Receiver $1$} &  $\mathbf{H}_{21}\mathbf{P}_{21}\preceq\mathbf{H}_{11}\mathbf{Q}_1$ & $\mathbf{H}_{11}\tilde{\mathbf{Q}}_1\preceq\mathbf{H}_{21}\mathbf{Q}_2$ & $\mathbf{H}_{21}\tilde{\mathbf{Q}}_2\preceq\mathbf{H}_{31}\mathbf{Q}_3$ & $\mathbf{H}_{21}\mathbf{P}_{24}\preceq\mathbf{H}_{31}\tilde{\mathbf{Q}}_3$\\
& $\mathbf{H}_{31}\mathbf{P}_{31}\preceq\mathbf{H}_{11}\mathbf{Q}_1$ & $\mathbf{H}_{31}\mathbf{P}_{32}\preceq\mathbf{H}_{21}\mathbf{Q}_2$ & & \\
\hline
\multirow{2}{*}{Receiver $2$} & & $\mathbf{H}_{12}\tilde{\mathbf{Q}}_1 \preceq \mathbf{H}_{22}\mathbf{Q}_2$ & $\mathbf{H}_{22}\tilde{\mathbf{Q}}_2 \preceq \mathbf{H}_{32}\mathbf{Q}_3$ &  \\
& $\mathbf{H}_{32}\mathbf{P}_{31}\preceq \mathbf{H}_{12}\mathbf{Q}_1$ & $\mathbf{H}_{32}\mathbf{P}_{32}\preceq \mathbf{H}_{22}\mathbf{Q}_2$ & $\mathbf{H}_{12}\mathbf{P}_{13} \preceq \mathbf{H}_{32}\mathbf{Q}_3$ & $\mathbf{H}_{12}\mathbf{P}_{14}\preceq\mathbf{H}_{32}\tilde{\mathbf{Q}}_3$  \\
\hline
\multirow{2}{*}{Receiver $3$} & $\mathbf{H}_{23}\mathbf{P}_{21}\preceq \mathbf{H}_{13}\mathbf{Q}_1$ & $\mathbf{H}_{13}\tilde{\mathbf{Q}}_1 \preceq \mathbf{H}_{23}\mathbf{Q}_2$ & $\mathbf{H}_{23}\tilde{\mathbf{Q}}_{2} \preceq \mathbf{H}_{33}\mathbf{Q}_3$ & $\mathbf{H}_{23}\mathbf{P}_{24}\preceq\mathbf{H}_{33}\tilde{\mathbf{Q}}_3$ \\
&&& $\mathbf{H}_{13}\mathbf{P}_{13} \preceq \mathbf{H}_{33}\mathbf{Q}_3$ & $\mathbf{H}_{13}\mathbf{P}_{14}\preceq\mathbf{H}_{33}\tilde{\mathbf{Q}}_3$\\
\hline
\end{tabular}
\caption{Summary of alignment equations.}
\label{table:align_eqns}
\end{table}

Now, me make the following selections:
\begin{align}
\mathbf{P}_{21}=\mathbf{P}_{31} \stackrel{\Delta}{=}&\tilde{\mathbf{P}}_1\\
\mathbf{P}_{32} \stackrel{\Delta}{=}& \tilde{\mathbf{P}}_2\\
\mathbf{P}_{13} \stackrel{\Delta}{=}& \tilde{\mathbf{P}}_3\\
\mathbf{P}_{14} = \mathbf{P}_{24}\stackrel{\Delta}{=}&\tilde{\mathbf{P}}_4\\
\tilde{\mathbf{Q}}_1 =& \mathbf{H}_{11}^{-1}\mathbf{H}_{31}\tilde{\mathbf{P}}_2\label{eq:q1_align}\\
\tilde{\mathbf{Q}}_2 =& \mathbf{H}_{22}^{-1}\mathbf{H}_{12}\tilde{\mathbf{P}}_3 \label{eq:q2_align}
\end{align}
Note that \eqref{eq:q1_align} and \eqref{eq:q2_align} imply that the artificial noises $\tilde{\mathbf{u}}_{1}$ and $\tilde{\mathbf{u}}_{2}$ align exactly with unintended message symbols $\mathbf{v}_{32}$ and $\mathbf{v}_{13}$ at receivers $1$ and $2$, respectively. With these selections, it suffices to find matrices $\tilde{\mathbf{P}}_i, i=1,\ldots,4$, $\mathbf{Q}_i, i=1,2,3$, and $\tilde{\mathbf{Q}}_3$. The alignment equations may now be written as
\begin{align}
\mathbf{T}_{ij}\tilde{\mathbf{P}}_i \preceq& \mathbf{Q}_i,\quad i=1,2,3, \quad j=1,\ldots,4\\
\mathbf{T}_{4j}\tilde{\mathbf{P}}_4 \preceq& \tilde{\mathbf{Q}}_3, \quad j=1,\ldots,4
\end{align}
where the $T_{ij}$s are tabulated in Table~\ref{table:final_align}, and the notation $\mathbf{A} \preceq \mathbf{B}$ is used to denote that $CS(\mathbf{A}) \subseteq CS(\mathbf{B})$ for matrices $\mathbf{A}$ and $\mathbf{B}$ where $CS(\mathbf{X})$ refers to the column space of the matrix $\mathbf{X}$.

\begin{table}[t]
\renewcommand{\arraystretch}{1.5}
\centering
\begin{tabular}{|c|c|c|c|c|}
\hline
& $T_{1j}$ & $T_{2j}$ & $T_{3j}$ & $T_{4j}$\\
\hline
$j=1$ & $\mathbf{H}_{11}^{-1}\mathbf{H}_{21}$ & $\mathbf{H}_{21}^{-1}\mathbf{H}_{31}$ & $\mathbf{H}_{31}^{-1}\mathbf{H}_{21}\mathbf{H}_{22}^{-1}\mathbf{H}_{12}$ & $\mathbf{H}_{31}^{-1}\mathbf{H}_{21}$\\
\hline
$j=2$ & $\mathbf{H}_{11}^{-1}\mathbf{H}_{31}$ & $\mathbf{H}_{22}^{-1}\mathbf{H}_{12}\mathbf{H}_{11}^{-1}\mathbf{H}_{31}$ & $\mathbf{H}_{32}^{-1}\mathbf{H}_{12}$ & $\mathbf{H}_{32}^{-1}\mathbf{H}_{12}$ \\
\hline
$j=3$ & $\mathbf{H}_{12}^{-1}\mathbf{H}_{32}$ & $\mathbf{H}_{22}^{-1}\mathbf{H}_{32}$ & $\mathbf{H}_{33}^{-1}\mathbf{H}_{23}\mathbf{H}_{22}^{-1}\mathbf{H}_{12}$ & $\mathbf{H}_{33}^{-1}\mathbf{H}_{23}$ \\
\hline
$j=4$ & $\mathbf{H}_{13}^{-1}\mathbf{H}_{23}$ & $\mathbf{H}_{23}^{-1}\mathbf{H}_{13}\mathbf{H}_{11}^{-1}\mathbf{H}_{31}$ & $\mathbf{H}_{33}^{-1}\mathbf{H}_{13}$ & $\mathbf{H}_{33}^{-1}\mathbf{H}_{13}$\\
\hline
\end{tabular}
\caption{Values of $T_{ij}$.}
\label{table:final_align}
\end{table}

We can now construct the matrices $\tilde{\mathbf{P}}_i, i=1,\ldots,4$, $\mathbf{Q}_i, i=1,\ldots,3$ and $\tilde{\mathbf{Q}}_{3}$ as in \cite{cadambe_jafar_interference}
\begin{align}
\tilde{\mathbf{P}}_i =& \left\lbrace \left( \prod\limits_{j=1}^{4} \mathbf{T}_{ij}^{\alpha_{j}}\right)\mathbf{w}_i : \alpha_{j} \in \left\lbrace 1,\ldots,n\right\rbrace   \right\rbrace \\
{\mathbf{Q}}_i =& \left\lbrace \left( \prod\limits_{j=1}^4 \mathbf{T}_{ij}^{\alpha_j}\right)\mathbf{w}_i : \alpha_{j} \in \left\lbrace 1,\ldots,n+1\right\rbrace   \right\rbrace\\
\tilde{\mathbf{Q}}_3 =& \left\lbrace \left( \prod\limits_{j=1}^4 \mathbf{T}_{ij}^{\alpha_j}\right)\mathbf{w}_{4} : \alpha_{j} \in \left\lbrace 1,\ldots,n+1\right\rbrace   \right\rbrace
\end{align}
where each $\mathbf{w}_i$ is the $M_n\times 1$ column vector containing elements drawn independently from a continuous distribution with bounded support. Note that an element in $\mathbf{P}_i$ is the product of powers of some channel coefficients and an extra random variable, just like an element in the sets $T_i$ defined for the real interference scheme. Further, the set of channel coefficients appearing in $\mathbf{P}_i$ is the same as those contained in set $T_i$. Thus, there is a loose correspondence between the real and vector space alignment techniques.

Now, consider the decodability of the desired signals at the receivers. For example, consider receiver $1$. Due to the alignment conditions in Table \ref{table:align_eqns}, the interference subspace at receiver $1$ is given by
\begin{align}
\mathbf{I}_1 = \left[ \mathbf{H}_{11}\mathbf{Q}_1\quad \mathbf{H}_{21}\mathbf{Q}_2\quad \mathbf{H}_{31}\mathbf{Q}_3 \quad \mathbf{H_{31}}\tilde{\mathbf{Q}}_3\right]
\end{align}
The desired signal subspace, on the other hand, is
\begin{align}
\mathbf{D}_1 = \left[\mathbf{H}_{11}\tilde{\mathbf{P}}_3\quad \mathbf{H}_{11}\tilde{\mathbf{P}}_4 \right]
\end{align}
For decodability, it suffices to show that
\begin{align}
\mathbf{\Lambda}_1 = \left[\mathbf{D}_1\quad \mathbf{I}_1 \right]
\end{align}
is full rank. To do so, we use \cite[Lemmas 1, 2]{cadambe_jafar_x_channel}. Consider any row $m$ of the matrix $\mathbf{\Lambda}_1$. Note that the $m$th row of $\mathbf{H}_{i1}\mathbf{Q}_i$ contains the term $w_{mi}$ with exponent $1$, but no $w_{mj}$ for $i\neq j$, where $w_{mi}$ denotes the element in the $m$th row of $\mathbf{w}_i$. In fact, for $i=1,\ldots,4$, the term $w_{mi}$ occurs nowhere else in the matrix $\mathbf{\Lambda}_l$ except in $\mathbf{H}_{i1}\mathbf{Q}_i$ ($\mathbf{H}_{31}\tilde{\mathbf{Q}}_3$, when $i=4$) and $\mathbf{H}_{11}\tilde{\mathbf{P}}_i$. This shows that $\mathbf{D}_1$ and $\mathbf{I}_1$ have full column ranks individually. Further, the matrix $\left[\mathbf{H}_{11}\tilde{\mathbf{P}}_3\quad \mathbf{H}_{31}\mathbf{Q}_3 \right] $ has full column rank because $\mathbf{Q}_3$ does not contain any elements of $\mathbf{H}_{11}$. Similarly, $\left[\mathbf{H}_{11}\tilde{\mathbf{P}}_4\quad {\mathbf{H}}_{31}\tilde{\mathbf{Q}}_3 \right] $ is full column rank for the same reason. Thus, $\mathbf{\Lambda}_1$, which is a $M_n \times M_n$ matrix, is full column rank, and hence full rank. This ensures decodability of the desired signals at receiver $1$. a similar analysis holds for the other receivers as well.

The security of the message signals at the eavesdropper is ensured by the fact that the artificial noises $\mathbf{Q}_i$ and $\tilde{\mathbf{Q}}_i$, $i=1,2,3$, do not align at the eavesdropper, and instead span the full received signal space at the eavesdropper. Indeed, the $M_n\times M_n$ matrix
\begin{align}
\mathbf{I}_E = \left[ \mathbf{G}_1\mathbf{Q}_1\quad \mathbf{G}_2\mathbf{Q}_2 \quad \mathbf{G}_3\mathbf{Q}_3\quad \mathbf{G}_1\tilde{\mathbf{Q}}_1\quad \mathbf{G}_2\tilde{\mathbf{Q}}_2 \quad \mathbf{G}_3\tilde{\mathbf{Q}}_3\right]
\end{align}
is full rank. Thus, if $\mathbf{V}_i = \left\lbrace  \mathbf{v}_{ij}, j\neq i,i+1 \right\rbrace$ denotes the collection of all messages of transmitter $i$, and $\mathbf{u}^T= \left[\mathbf{u}_1^T,\mathbf{u}_2^T,\mathbf{u}_2^T,\tilde{\mathbf{u}}_1^T,\tilde{\mathbf{u}}_2^T,\tilde{\mathbf{u}}_3^T   \right] $,
\begin{align}
I(\mathbf{V}_1^3;\mathbf{Z}) =& h(\mathbf{Z}) - h(\mathbf{Z}|\mathbf{V}_1^3)\\
=& h(\mathbf{Z}) - h(\mathbf{I}_E\mathbf{u})\\
\leq& \frac{M_n}{2}\log P - \frac{M_n}{2}\log P +o(\log P)\\
=& o(\log P)
\end{align}
In the above calculation, we have dropped the conditioning on $\Omega$ for notational simplicity.
Now, by treating all $M_n$ channel uses as $1$ vector channel use, and using \cite[Theorem~2]{jianwei_ulukus_interference_2013}, an achievable rate for the vector channel is
\begin{align}
R_i^{M_n} =& I(\mathbf{V}_i;\mathbf{Y}_i) - I(\mathbf{V}_i;\mathbf{Z}|\mathbf{V}_{-i})\\
=& 2n^{\Gamma}\log P - o(\log P) \label{eq:intermed_step}
\end{align}
where \eqref{eq:intermed_step} follows since the $2n^{\Gamma}$ symbols are decodable within noise variance, and since $I(\mathbf{V}_i;\mathbf{Z}|\mathbf{V}_{-i}) \leq I(\mathbf{V}_1^3;\mathbf{Z})\leq o(\log P)$.
Thus, the rate $\frac{2n^{\Gamma}}{M_n}$ is achievable per user pair per channel use, which gives a sum s.d.o.f.~of $\frac{6n^{\Gamma}}{2n^{\Gamma}+ 4(n+1)^{\Gamma}}$, which approaches $1$, as $n\rightarrow \infty$.

We remark here that our scheme also provides confidentiality, that is, the messages from transmitter $i$ are kept secure from legitimate receiver $j$. We get this confidentiality without any additional loss of rate, just as in the case when eavesdropper CSI is available at the transmitters \cite{jianwei_ulukus_interference_2013}.

\subsection{Converse for the Interference Channel}

The steps of the converse are similar to that of the proof in Section~\ref{converse:mac}. The notation here is also the same as in Section~\ref{converse:mac}. Again, we divide the proof into three steps.

\subsubsection{Deterministic Channel  Model}

We consider the deterministic channel given as,
\begin{align}
Y_k(t) =& \sum\limits_{i=1}^{K} \lf h_{ik}(t)X_i(t) \rf\label{eq:inter_detmod1}\\
Z(t) =& \sum\limits_{i=1}^{K}\lf g_i(t)X_i (t)\rf\label{eq:inter_detmod2}
\end{align}
for $k=1,\ldots,K$, with the constraint that
\begin{align}
X_{i}(t) \in \left\{ 0,1,\ldots, \lf \sqrt{P} \rf \right\}\label{eq:inter_pow_constraints}
\end{align}
We can show that there is no loss of s.d.o.f.~in considering the channel in \eqref{eq:inter_detmod1}-\eqref{eq:inter_detmod2} instead of the one in \eqref{eq:inter_model1}-\eqref{eq:inter_model2}, as in Section~\ref{sec:det_model}.
Thus, we will consider this deterministic channel in the remaining part of the converse. Since all receivers know $\Omega$, it appears in the conditioning in every entropy and mutual information term below. We keep this in mind, but drop it for the sake of notational simplicity.

\subsubsection{An Upper Bound on the Sum Rate}

We begin as in the \emph{secrecy penalty lemma} in \cite{jianwei_ulukus_one_hop}, i.e., \cite[Lemma~1]{jianwei_ulukus_one_hop}. Note that, unlike \cite[Lemma~1]{jianwei_ulukus_one_hop}, channel inputs are integer here:
\begin{align}
n \sum_{i=1}^{K}R_i \leq& I(W_1^K;\mathbf{Y}_1^K) - I(W_1^K;\mathbf{Z}) +n\epsilon\\
\leq& I(W_1^K;\mathbf{Y}_1^K|\mathbf{Z}) +n\epsilon\\
\leq& I(\mathbf{X}_1^K;\mathbf{Y}_1^K|\mathbf{Z}) +n\epsilon\\
\leq& H(\mathbf{Y}_1^K|\mathbf{Z}) +n\epsilon\label{eq:inter_dep1}\\
=& H(\mathbf{Y}_1^K,\mathbf{Z}) - H(\mathbf{Z}) +n\epsilon\\
\leq& H(\mathbf{X}_1^K, \mathbf{Y}_1^K,\mathbf{Z})- H(\mathbf{Z}) +n\epsilon\\
=& H(\mathbf{X}_1^K)- H(\mathbf{Z}) +n\epsilon\label{eq:inter_dep}\\
\leq& \sum\limits_{k=1}^{K} H(\mathbf{X}_k) -H(\mathbf{Z})  +n\epsilon\label{eq:inter_inter}
\end{align}
where \eqref{eq:inter_dep} follows since $H(\mathbf{Y}_1^K,\mathbf{Z}|\mathbf{X}_1^K)=0$ for the channel in \eqref{eq:inter_detmod1}-\eqref{eq:inter_detmod2}.

Also, to ensure decodability at the legitimate receiver, we use the \emph{role of a helper lemma} in \cite{jianwei_ulukus_one_hop}, i.e., \cite[Lemma~2]{jianwei_ulukus_one_hop},
\begin{align}
n R_i \leq& I(W_i;\mathbf{Y}_i) +n\epsilon'\\
\leq& I(\mathbf{X}_i;\mathbf{Y}_i)+n\epsilon'\\
=& H(\mathbf{Y}_i) - H(\mathbf{Y}_i|\mathbf{X}_i)+n\epsilon'\\
=& H(\mathbf{Y}_i) - H(\lf \mathbf{h}_j\mathbf{X}_j\rf)+n\epsilon' \\
=& H(\mathbf{Y}_i) - H(\lf \mathbf{h}_j\mathbf{X}_j\rf, \mathbf{X}_j)+ H(\mathbf{X}_j|\lf \mathbf{h}_j\mathbf{X}_j\rf)+n\epsilon'\\
\leq& H(\mathbf{Y}_i) - H(\mathbf{X}_j)+ H(\mathbf{X}_j|\lf \mathbf{h}_j\mathbf{X}_j\rf)+n\epsilon'\\
\leq& H(\mathbf{Y}_i) - H(\mathbf{X}_j)+ \sum_{t=1}^{n} H(X_j(t)|\lf {h}_j(t)X_j(t)\rf)+n\epsilon'\label{eq:inter_decode_step2}\\
\leq& H(\mathbf{Y}_i) - H(\mathbf{X}_j)+n\epsilon'+nc \label{eq:inter_decode1}
\end{align}
for every $i\neq j$, where \eqref{eq:inter_decode1} follows using Lemma~\ref{lem:inter_step}.

Let $\Pi$ be any derangement of $\left(1,\ldots,n\right) $, and let $j=\Pi(i)$. Then, using \eqref{eq:inter_decode1}, we obtain,
\begin{align}
\sum_{k=1}^K H(\mathbf{X}_k) \leq \sum_{k=1}^K H(\mathbf{Y}_k) - n\sum_{k=1}^K R_k +nK(\epsilon'+c)\label{eq:inter_step}
\end{align}
Using \eqref{eq:inter_step} in \eqref{eq:inter_inter}, we get,
\begin{align}
2n\sum\limits_{i=1}^{K} R_i \leq& \sum_{k=1}^K H(\mathbf{Y}_k) - H(\mathbf{Z})+nK(\epsilon'+c)+n\epsilon\\
\leq& (K-1)\frac{n}{2}\log P + \left(H(\mathbf{Y}_K) - H(\mathbf{Z})\right)+n\epsilon''
\end{align}
where $\epsilon'' =  o(\log P)$. Dividing by $n$ and letting $n\rightarrow \infty$,
\begin{align}
2 \sum\limits_{i=1}^{K} R_i\leq& (K-1)\frac{1}{2}\log P +\lim_{n\rightarrow \infty} \frac{1}{n}\left(H(\mathbf{Y}_K) - H(\mathbf{Z})\right)+  \epsilon''
\end{align}
Now dividing by $\frac{1}{2}\log P$ and taking $P\rightarrow \infty$,
\begin{align}
\sum\limits_{i=1}^{K} d_i \leq& \frac{K-1}{2} + \frac{1}{2}\lim_{P \rightarrow \infty}\lim_{n\rightarrow \infty} \frac{H(\mathbf{Y}_K) - H(\mathbf{Z})}{\frac{n}{2}\log P}\label{eq:inter_rate_bound}
\end{align}

\subsubsection{Bounding the Difference of Entropies}

As we did in Section~\ref{sec:mac_entropy_bound}, we enhance the system by relaxing the condition that channel inputs from different transmitters are mutually independent, and think of the $K$ single antenna terminals as a single transmitter with $K$ antennas. Thus, we wish to maximize $H(\mathbf{Y}_K) - H(\mathbf{Z})$, where $\mathbf{Y}_K$ and $\mathbf{Z}$ are two single antenna receiver outputs, under the constraint that the channel gains to $\mathbf{Z}$ are unknown at the transmitter. This again brings us to the $K$-user MISO broadcast channel setting of \cite{aligned_image_sets_jafar}. We know from \cite[eqns.~(75)-(103)]{aligned_image_sets_jafar} that even without any security or decodability constraints, the difference of entropies, $H(\mathbf{Y}_K) - H(\mathbf{Z})$ cannot be larger than $no(\log P)$, if the channel gains to the second receiver is unknown. Thus,
\begin{align}
H(\mathbf{Y}_K) - H(\mathbf{Z}) \leq no(\log P) \label{eq:inter_entropy_bound}
\end{align}

Using \eqref{eq:inter_entropy_bound} in \eqref{eq:inter_rate_bound}, we have
\begin{align}
\sum\limits_{i=1}^{K} d_i \leq& \frac{K-1}{2}
\end{align}
This completes the converse proof of Theorem~\ref{theo:interference}.

\section{Proof of Theorem~\ref{theo:m_mac}}\label{sec:theo_mac2_proof}

In this section, we present achievable schemes that achieve sum s.d.o.f.~of $\frac{m(K-1)}{m(K-1)+1}$, when $m$ of the $K$ transmitters have eavesdropper's CSI, for both fixed and fading channel gains.

\begin{figure}[t]
\centering
\includegraphics[width=\linewidth]{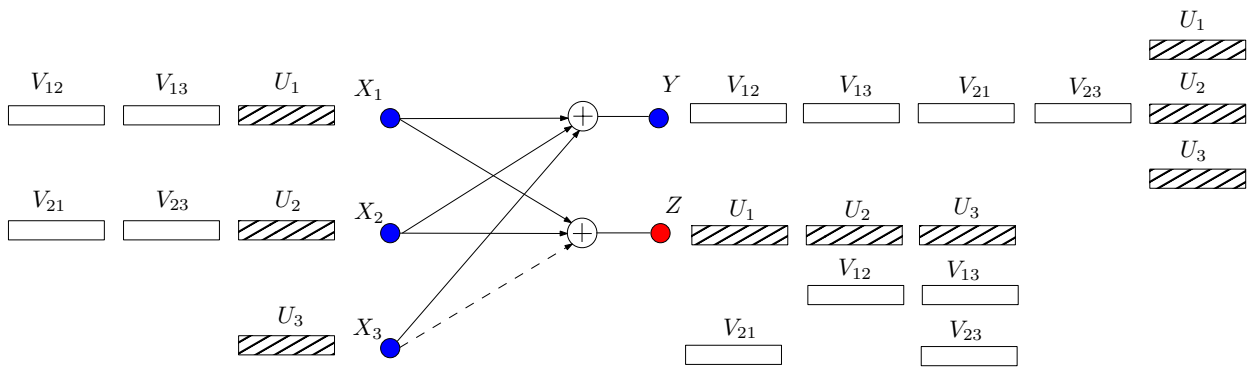}
\caption{Alignment of signals when $K=3$ and $m=2$.}
\label{fig:alignment_3_users_partial_csi}
\end{figure}

\subsection{Fixed Channel Gains}

With fixed channel gains, we provide a scheme based on real interference alignment that achieves the required sum s.d.o.f.~of $\frac{m(K-1)}{m(K-1)+1}$. In particular, it achieves the s.d.o.f.~tuple $\left(d_1,\ldots,d_m,d_{m+1},\ldots,d_K\right) = \left(\frac{K-1}{m(K-1)+1},\ldots,\frac{K-1}{m(K-1)+1},0,\ldots,0 \right)$. We employ $m(K-1)+K$ mutually independent random variables:
\begin{align}
V_{ij},&\quad i=1,\ldots,m, j=1,\ldots,K, j\neq i\nonumber\\
U_j,&\quad j=1,\ldots,K\nonumber
\end{align}
uniformly drawn from the same PAM constellation $C(a,Q)$ in \eqref{eq:constellation}. Transmitter $i$, $i=1,\ldots,m$ transmits:
\begin{align}
X_i=\sum\limits_{j=1, j\neq i}^{K} \frac{g_j}{h_jg_{i}}V_{ij} + \frac{1}{h_{i}}U_i, \quad i=1,\ldots,m
\end{align}
while transmitters $(m+1)$ to $K$ transmit
\begin{align}
X_i=\frac{1}{h_i}U_i, \quad i=m+1,\ldots,K
\end{align}
The channel outputs are given by,
\begin{align}
Y =& \sum_{i=1}^{m}\sum_{j\neq i} \frac{h_ig_j}{h_jg_{i}}V_{ij} + \sum_{i=1}^{K}U_i+N_1\\
Z =& \sum_{i=1}^{K} \frac{g_i}{h_i}\left(U_i + \sum_{j=1, j\neq i}^{m} V_{ji}\right)+N_2
\end{align}
Intuitively, every $V_{ij}$ gets superimposed with $U_j$ at the eavesdropper, thus securing it. This is shown in Fig.~\ref{fig:alignment_3_users_partial_csi}. The proof of decodability and security guarantee follows exactly the proof in \cite[Section~IX-B ]{jianwei_ulukus_one_hop} and is omitted here.

\subsection{Fading Channel Gains}

We construct a scheme that achieves the desired sum s.d.o.f. Without loss of generality, assume that the first $m$ transmitters have eavesdropper CSI, while the remaining transmitters have no eavesdropper CSI. We provide a scheme to achieve the rate tuple $\left(d_1,\ldots,d_m,d_{m+1},\right. \\ \left.\ldots,d_K\vphantom{d_K}\right) = \left(\frac{K-1}{m(K-1)+1},\ldots,\frac{K-1}{m(K-1)+1},0,\ldots,0 \right)$, thus, achieving the required sum s.d.o.f.~of $\frac{m(K-1)}{m(K-1)+1}$. For each $i=1,\ldots, m$, transmitter $i$ sends $\mathbf{V}_i = \left\{ V_{ij}, , j\neq i, j=1,\ldots,K\right\}$ symbols in $m(K-1)+1$ time slots. Let $\mathbf{V} = \left\{\mathbf{V}_i, i=1,\ldots,K\right\}$. Fig.~\ref{fig:alignment_3_users_partial_csi} illustrates the alignment of the signals at the end of the scheme when $K=3$ and $m=2$. The scheme is as follows:

At time $t\in \left\{ 1,\ldots,m(K-1)+1\right\}$, the $i$th transmitter, $i=1,\ldots, K$, sends,
\begin{align}
X_i(t) = \begin{cases}
\sum\limits_{j=1, j\neq i}^{K} \frac{g_j(t)}{h_j(t)g_{i}(t)}V_{ij} + \frac{1}{h_{i}(t)}U_i,& 1\leq i \leq m\\
\frac{1}{h_i(t)}U_i,& m+1\leq i\leq K
\end{cases}
\end{align}
where $U_i$ is an artificial noise symbol. This ensures that the noise symbols $U_i$ all align at the legitimate receiver. On the other hand, the artificial noise symbol from the $j$th transmitter $U_j$ protects all the messages $V_{ij}$ for every $i$, at the eavesdropper. The channel outputs are given by,
\begin{align}
Y(t) =& \sum_{i=1}^{m}\sum_{j\neq i} \frac{h_i(t)g_j(t)}{h_j(t)g_{i}(t)}V_{ij} + \sum_{i=1}^{K}U_i+N_1(t)\\
Z(t) =& \sum_{i=1}^{K} \frac{g_i(t)}{h_i(t)}\left(U_i + \sum_{j=1, j\neq i}^{m} V_{ji}\right)+N_2(t)
\end{align}

After the $m(K-1)+1$ time slots, the legitimate receiver ends up with $m(K-1)+1$ linearly independent equations with $m(K-1)+1$ variables: $\sum_{i=1}^{K}U_i$ and the $m(K-1)$ variables $ \left\lbrace V_{ij}\right\rbrace $. Thus, it can decode all the $m(K-1)$ message symbols $V_{ij}$. Defining $\mathbf{Y} = \left\{ Y(t), ~t=1,\ldots, m(K-1)+1\right\}$ and $\mathbf{Z}$ similarly as $\mathbf{Y}$, this means that $I(\mathbf{V};\mathbf{Y}) =m(K-1)\frac{1}{2}\log P + o(\log P)$, and also $I(\mathbf{V};\mathbf{Z})\leq o(\log P)$, concluding the achievability proof.

\section{Conclusions}

In this paper, we established the optimal sum s.d.o.f.~for three channel models: the wiretap channel with $M$ helpers, the $K$-user multiple access wiretap channel, and the $K$-user interference channel with an external eavesdropper, in the absence of eavesdropper's CSIT. While there is no loss in the s.d.o.f.~for the wiretap channel with helpers in the absence of the eavesdropper's CSIT, the s.d.o.f.~decreases in the cases of the multiple access wiretap channel and the interference channel with an external eavesdropper. We show that in the absence of eavesdropper's CSIT, the $K$-user multiple access wiretap channel is equivalent to a wiretap channel with $(K-1)$ helpers from a sum s.d.o.f.~perspective. The question of optimality of the sum s.d.o.f.~when some but not all of the transmitters have the eavesdropper's CSIT remains a subject of future work.

\begin{appendices}

\section{Proof of Lemma~\ref{lem:dif_entropy}} \label{a:dif_entropy}

Since $\mathbf{A}\mathbf{X} + \mathbf{N}$ is a jointly Gaussian random vector with zero-mean and covariance $P\mathbf{A}\mathbf{A}^T + \sigma^2\mathbf{I}$, we have \cite{cover_book},
\begin{align}
h(\mathbf{A}\mathbf{X} + \mathbf{N}) =&\frac{1}{2} \log(2\pi e)^M \left|P\mathbf{A}\mathbf{A}^T + \sigma^2\mathbf{I}\right|\\
=& \frac{1}{2} \log(2\pi e)^M \left|P\mathbf{W\Sigma W}^T + \sigma^2\mathbf{I}\right|\\
=& \frac{1}{2} \sum_{i=1}^r \log \left(\lambda_iP+\sigma^2\right) + o(\log P)\\
=& r \left(\frac{1}{2}\log P\right) +o(\log P)
\end{align}
where we note that $\mathbf{A}\mathbf{A}^T$ is positive semi-definite, with an eigenvalue decomposition $\mathbf{W}\mathbf{\Sigma} \mathbf{W}^T$, where $\mathbf{\Sigma}$ is a diagonal matrix with $r$ non-zero entries $\lambda_1,\ldots,\lambda_r$.

\section{Proof of Lemma~\ref{lem:inter_step}} \label{a:proof_of_lemma_inter_step}

First, note that
\begin{align}
H(X|\lf hX\rf,h)= E_h\left[H(X|\lf hX \rf, h=\tilde{h})\right] \label{eq:decode_step4}
\end{align}
Now, for a fixed $h$, let us define $S_{h}(\nu)$ as the set of all realizations of $X$ such that $\lf hX\rf = \nu$, i.e., $S_{h}(\nu) \stackrel{\Delta}{=} \left\lbrace i\in \left\lbrace 1,\ldots,\lfloor\sqrt{P}\rfloor\right\rbrace : \lf ih\rf =\nu  \right\rbrace $. Then,
\begin{align}
H\left(X|\lf hX\rf,h=\tilde{h}\right) \leq \log |S_{\tilde{h}}(\lfloor \tilde{h}X\rfloor)| \label{eq:decode_step5}
\end{align}
For any $\nu$, we can upper-bound $|S_{\tilde{h}}(\nu)|$ as follows: Let, $i_1$ and $i_2$ be the minimum and maximum elements of $S_{\tilde{h}}(\nu)$. Then, $\lfloor i_1\tilde{h}\rfloor = \lfloor i_2\tilde{h}\rfloor$ implies that $(i_2-i_1)|\tilde{h}|<1$, which means $(i_2-i_1) < \frac{1}{|\tilde{h}|}$. Hence,
\begin{align}
|S_{\tilde{h}}(\nu)| \leq& i_2-i_1 +1\\
<& 1+ \frac{1}{|\tilde{h}|}
\end{align}
Thus, using \eqref{eq:decode_step4} and \eqref{eq:decode_step5}, we have,
\begin{align}
H\left(X|\lf {h}X\rf,h\right) \leq& E_{h}\left[\log\left(1+\frac{1}{|h|}\right)\right]\leq c \label{eq:decode_step6}
\end{align}
where $c$ is a constant independent of $P$.

\section{Achievability for the $K$-user Interference Channel with an External Eavesdropper}\label{k_user_schemes}

Here, we present the general achievable schemes for the $K$-user interference channel with an external eavesdropper.     
\subsection{Fixed Channel Gains}\label{a:fixed}

Let $m$ be a large constant. We pick $(K+1)$ points $c_1,\ldots,c_{K+1}$ in an i.i.d.~fashion from a continuous distribution with bounded support. Then, $c_1,\ldots,c_{K+1}$ are rationally independent almost surely. Let us define sets $T_i$, for $i=1,\ldots,K+1$, which will represent \emph{dimensions}
as follows:
\begin{align}
T_1 \stackrel{\Delta}{=}& \left\{
      \left( \prod_{k=1}^{K}h_{1k}^{r_{1k}}\right)
      \left(
        \prod_{j,k=1, j\ne 1,k}^{K} h_{jk}^{r_{jk}}
      \right)c_1^{s}
      :
      ~r_{jk}, s \in \{1,\ldots,m\}
    \right\}
    \\
T_{i}\stackrel{\Delta}{=}& \left\{
      \left(\prod_{k=1}^{K}h_{ik}^{r_{ik}}\right)
      \left(\prod_{k=2}^{K}\left(\frac{h_{(i-1)k}}{h_{(i-1)1}}\right)^{r_{(i-1)k}} \right)
            \left(
              \prod_{\substack{j,k=1\\ j\ne i,i-1,k}}^{K} h_{jk}^{r_{jk}}
            \right)c_i^{s}
            :
            ~r_{jk}, s \in \{1,\ldots,m\}
          \right\}, \nonumber\\ &\hspace{300 pt} i=2,\ldots,K-1\\
T_{K}\stackrel{\Delta}{=}& \left\{
      \left(\prod_{k=1}^{K}h_{Kk}^{r_{Kk}}\right)
      \left(\prod_{k=1, k\neq 2}^{K} \left( \frac{h_{(K-1)k}}{h_{(K-1)2}}\right) ^{r_{(K-1)k}} \right)
            \left(
              \prod_{\substack{j,k=1\\ j\ne K,K-1,k}}^{K} h_{jk}^{r_{jk}}
            \right)c_K^{s}
            :
            ~r_{jk}, s \in \{1,\ldots,m\}
          \right\}\\
T_{K+1} \stackrel{\Delta}{=}& \left\{
      \left( \prod_{k=1}^{K}h_{Kk}^{r_{Kk}}\right)
      \left(
        \prod_{j,k=1, j\ne K,k}^{K} h_{jk}^{r_{jk}}
      \right)c_{K+1}^{s}
      :
      ~r_{jk}, s \in \{1,\ldots,m\}
    \right\}
\end{align}
Let $M_i$ be the cardinality of $T_i$. Note that all $M_i$ are the same, thus we denote them as $M$,
\begin{equation}
  M \stackrel{\Delta}{=} m^{2+K(K-1)} \label{eq:cardinality_real_ic}
 \end{equation}
First, we divide each message into many sub-messages; specifically, the message of the $i$th transmitter, $W_i$, is divided into $(K-1)$ sub-messages $V_{ij}, j=1,\ldots,K+1, j\neq i,i+1$. For each transmitter $i$, let $\mathbf{p}_{ij}$ be the vector containing all the elements of $T_j$, for $j\neq i,i+1$. For any given $(i,j)$ with $j\neq i,i+1$, $\mathbf{p}_{ij}$ represents the dimension along which message $V_{ij}$ is sent. Further, at each transmitter $i$, let $\mathbf{q}_i$ and $\tilde{\mathbf{q}}_{i}$ be vectors containing all the elements in sets $T_{i}$ and $\beta_iT_{i+1}$, respectively, where
\begin{align}
\beta_i = \begin{cases}
\frac{h_{(i+2)1}}{h_{i1}},\quad &\mbox{if } 1\leq i\leq K-2\\
\frac{h_{12}}{h_{i2}}, \quad &\mbox{if } i=K-1\\
1, &\mbox{if } i=K
\end{cases}
\end{align}
The vectors $\mathbf{q}_i$ and $\tilde{\mathbf{q}}_{i}$ represent dimensions along which artificial noise symbols $U_i$ and $\tilde{U}_i$, respectively, are sent. We define a $(K+1)M$ dimensional vector $\mathbf{b}_i$ by stacking the $\mathbf{p}_{ij}$s, $\mathbf{q}_i$ and $\tilde{\mathbf{q}}_i$ as
\begin{align}
\mathbf{b}_i^T = \left[ \mathbf{p}_{i1}^T \ldots \mathbf{p}_{i(i-1)}^T \quad \mathbf{p}_{i(i+2)}^T \ldots \mathbf{p}_{i(K+1)}\quad \mathbf{q}_i \quad \tilde{\mathbf{q}}_i \right]
\end{align}
The transmitter encodes $V_{ij}$ using an $M$ dimensional vector $\mathbf{v}_{ij}$, and the cooperative jamming signals $U_i$ and $\tilde{U}_i$ using $M$ dimensional vectors $\mathbf{u}_i$ and $\tilde{\mathbf{u}}_i$, respectively. Each element of $\mathbf{v}_{ij}$, $\mathbf{u}_i$ and $\tilde{\mathbf{u}}_i$ are drawn in an i.i.d.~fashion from $C(a,Q)$ in \eqref{eq:constellation}. Let
\begin{align}
\mathbf{a}_i^T =\left[ \mathbf{v}_{i1}^T \ldots \mathbf{v}_{i(i-1)}^T \quad \mathbf{v}_{i(i+2)}^T \ldots \mathbf{v}_{i(K+1)}\quad \mathbf{u}_i \quad \tilde{\mathbf{u}}_i \right]
\end{align}
The channel input of transmitter $i$ is then given by
\begin{align}
x_i = \mathbf{a}_i^T\mathbf{b}
\end{align}

\begin{figure}[t]
\centering
\includegraphics[width=\linewidth]{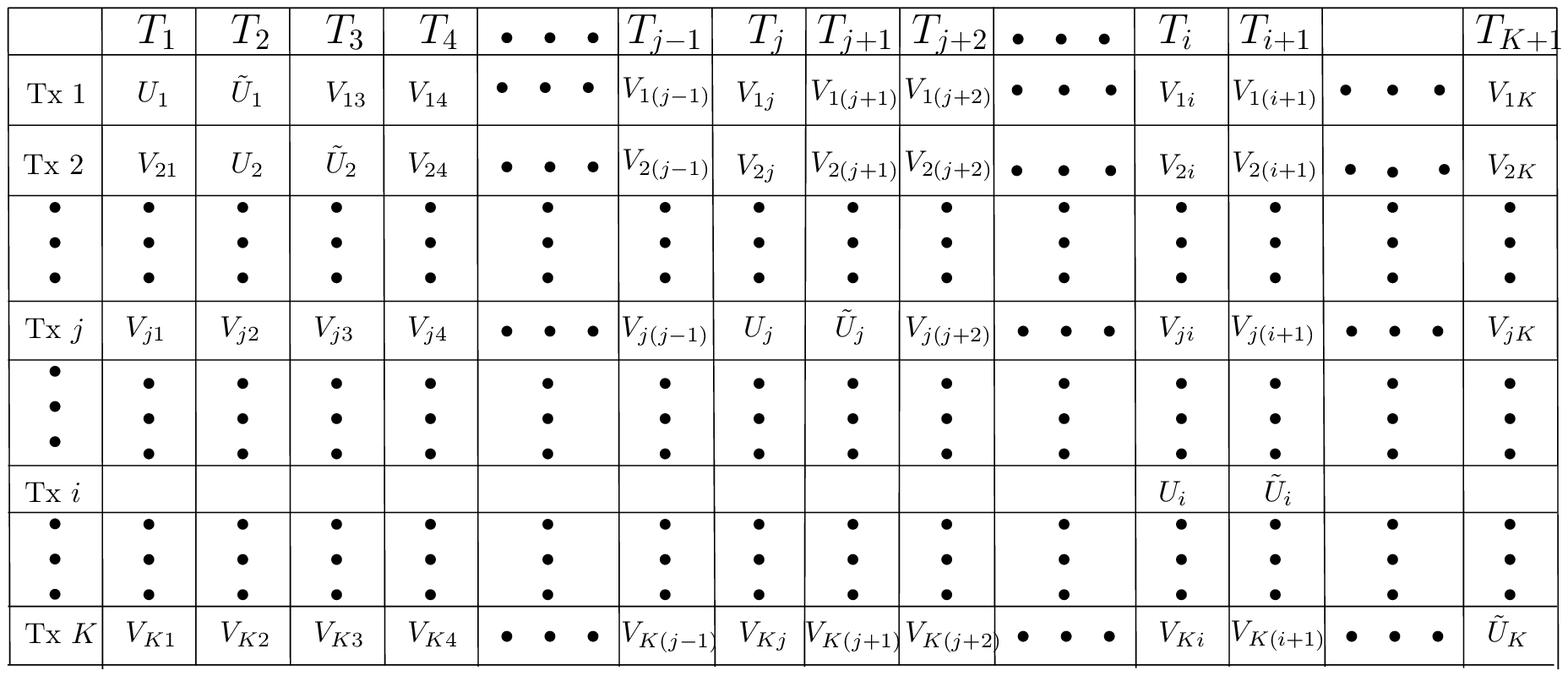}
\caption{Alignment of interference signals at receiver $i$.}
\label{fig:alignment_k_users}
\end{figure}

Let us now analyze the structure of the received signals at the legitimate receivers. The alignment of the interfering signal spaces at receiver $i$ is shown in Fig.~\ref{fig:alignment_k_users}. The $i$th row depicts the signals originating from transmitter $i$. The signals in the same column align together at the receiver. For simplicity of exposition, let us consider receiver $1$.

At the first receiver, the desired signals $\mathbf{v}_{13},\ldots$, $\mathbf{v}_{1(K+1)}$ come along dimensions $h_{11}T_3, \ldots$, $ h_{11}T_{K+1}$, respectively. These dimensions are \emph{separate} almost surely, since $T_i$ contains powers of $c_i$ while $T_j, j\neq i$ does not. Thus, they correspond to \emph{separate} boxes in the Fig.~\ref{fig:alignment_inter} for $K=3$. For the same reason, cooperative jamming signals $\mathbf{u}_1,\ldots$, $\mathbf{u}_K$, $\tilde{\mathbf{u}}_{K}$, which arrive along the dimensions $h_{11}T_1,\ldots$, $h_{K1}T_K$, $h_{K1}T_{K+1}$ occupy different dimensions almost surely. Further, the message signals $\mathbf{v}_{13},\ldots,\mathbf{v}_{1(K+1)}$, and the cooperative jamming signals $\mathbf{u}_1,\ldots,\mathbf{u}_K,\tilde{\mathbf{u}}_{K}$ do not overlap, since none of $T_3\ldots,T_{K+1}$ contain $h_{11}$. Thus, they appear as separate boxes in Fig.~\ref{fig:alignment_inter}.

Now, let us consider the signals that are not desired at receiver $1$. A signal $\mathbf{v}_{kl}, k\neq 1, K+1$ arrives at receiver $1$ along $h_{k1}T_l$. If we define
\begin{align}
\tilde{T}_1 \stackrel{\Delta}{=}& \left\{
      \left( \prod_{k=1}^{K}h_{1k}^{r_{1k}}\right)
      \left(
        \prod_{j,k=1, j\ne 1,k}^{K} h_{jk}^{r_{jk}}
      \right)c_1^{s}
      :
      ~r_{jk}, s \in \{1,\ldots,m+1\}
    \right\}
    \\
\tilde{T}_{i}\stackrel{\Delta}{=}& \left\{
      \left(\prod_{k=1}^{K}h_{ik}^{r_{ik}}\right)
      \left(\prod_{k=2}^{K}\left( \frac{h_{(i-1)k}}{h_{(i-1)1}}\right) ^{r_{(i-1)k}} \right)
            \left(
              \prod_{\substack{j,k=1\\ j\ne i,i-1,k}}^{K} h_{jk}^{r_{jk}}
            \right)c_i^{s}
            :
            ~r_{jk}, s \in \{1,\ldots,m+1\}
          \right\}, \nonumber\\ &\hspace{300 pt} i=2,\ldots,K-1\\
\tilde{T}_{K}\stackrel{\Delta}{=}& \left\{
      \left(\prod_{k=1}^{K}h_{Kk}^{r_{Kk}}\right)
      \left(\prod_{\substack{k=1, k\neq 2\\m=K-1}}^{K}\left( \frac{h_{mk}}{h_{m2}}\right) ^{r_{mk}} \right)
            \left(
              \prod_{\substack{j,k=1\\ j\ne K,K-1,k}}^{K} h_{jk}^{r_{jk}}
            \right)c_K^{s}
            :
            ~r_{jk}, s \in \{1,\ldots,m+1\}
          \right\}\\
\tilde{T}_{K+1} \stackrel{\Delta}{=}& \left\{
      \left( \prod_{k=1}^{K}h_{Kk}^{r_{Kk}}\right)
      \left(
        \prod_{j,k=1, j\ne K,k}^{K} h_{jk}^{r_{jk}}
      \right)c_{K+1}^{s}
      :
      ~r_{jk}, s \in \{1,\ldots,m+1\}
    \right\}
\end{align}
we notice that the dimensions in $h_{k1}T_l,\, k\neq 1$ are subsets of $\tilde{T}_l$, as is $h_{l1}T_l$ for every $l=1,\ldots,K$. Thus, each $\mathbf{v}_{kl}$ aligns with $\mathbf{u}_{l}$ in $\tilde{T}_l$, for $l=1,\ldots,K$, as is shown in Fig.~\ref{fig:alignment_k_users}. Further, a signal $\mathbf{v}_{k(K+1)}$, $k\neq 1,K$, arrives along the dimensions $h_{k1}T_{K+1}$, $k\neq 1$ which is a subset of $\tilde{T}_{K+1}$, as is $h_{K1}T_{K+1}$, along which $\tilde{\mathbf{u}}_K$ arrives. Thus, each $\mathbf{v}_{k(K+1)}$, $k\neq 1,K$ aligns with $\tilde{\mathbf{u}}_K$, see Fig.~\ref{fig:alignment_k_users}. Finally, the cooperative jamming signals $\tilde{\mathbf{u}}_1,\ldots,$ $\tilde{\mathbf{u}}_{K-2}$, and $\tilde{\mathbf{u}}_{K-1}$ arrive at receiver $1$ along dimensions $h_{31}T_2,\ldots$, $h_{K1}T_{K-1}$, and $h_{12}\left( \frac{h_{(K-1)1}}{h_{(K-1)2}}\right) T_K$, respectively, which are all in $\tilde{T}_2\ldots$, $\tilde{T}_{K-1}$ and $\tilde{T}_K$, respectively. Thus, the signal $\tilde{\mathbf{u}}_i, i=1,\ldots,K-1$ align with $\mathbf{u}_{i+1}$ in $\tilde{T}_{i+1}$, which is seen in Fig.~\ref{fig:alignment_inter} for $K=3$, and in Fig.~\ref{fig:alignment_k_users} for general $K$.

We further note that the sets $h_{11}T_3,\ldots$, $h_{11}T_{K+1}$, $\tilde{T}_1, \ldots$, $\tilde{T}_{K+1}$ are all separable since only $T_i$ and $\tilde{T}_i$ (and not $T_j$ or $\tilde{T_j}$) contain powers of $c_i$, and none of $\tilde{T_3},\ldots$, $\tilde{T}_{K+1}$ contains $h_{11}$. A similar observation holds for the received signal at any of the remaining receivers. Thus, the set
\begin{align}
S = \left( \bigcup_{i=3}^{K+1} h_{11}T_i\right)\bigcup\left( \bigcup_{i=1}^{K+1} \tilde{T}_i \right)
\end{align}
has cardinality given by
\begin{align}
M_s = (K-1)m^{K(K-1)+2} + (K+1)(m+1)^{K(K-1)+2}
\end{align}

At the external eavesdropper, there is no alignment and the cooperative jamming signals occupy the full space, thereby exhausting the decoding capability of the eavesdropper. This secures all the messages at the external eavesdropper.

We next provide an analysis for the achievable sum rate. Since we have only one eavesdropper, we use \cite[Theorem~2]{jianwei_ulukus_interference_2013} and observe that the rate
\begin{align}
R_i = I(V_i;Y_i) - I(V_i;Z|V_{-i}) \label{eq:rate_ic_equation}
\end{align}
is achievable, where $V_i$ ia an auxiliary random variable satisfying  $V_i \rightarrow X_i \rightarrow Y,Z$, and $V_{-i}$ denotes the collection $\left\lbrace V_j, j\neq i\right\rbrace$. Note that since $\Omega$ is known at all the legitimate receivers and the eavesdropper, and since $\mathbf{V}_i$s are  chosen to be independent of $\Omega$, $\Omega$ should appear in the conditioning of each of the mutual information quantities in \eqref{eq:rate_ic_equation}. We keep this in mind, but drop it for the sake of notational simplicity.

First, we can upper bound the probability of error at each receiver. Let
\begin{align}
V_i \stackrel{\Delta}{=} \left(\mathbf{v}_{i1}\ldots\mathbf{v}_{i(i-1)}\quad \mathbf{v}_{i(i+2)}\ldots \mathbf{v}_{i(K+1)}  \right)
\end{align}
Then, for any $\delta>0$, there exists a positive constant
$\gamma$, which is independent of $P$, such that if we choose $Q =
 P^{\frac{1-\delta}{2(M_S+\delta)}}$ and $a = \frac{\gamma
P^{\frac{1}{2}}}{Q}$,
then for almost all channel gains the average power constraint is satisfied
and the probability of error is bounded by
\begin{equation}
 \mathrm{Pr}(V_i\neq \hat V_i) \le\exp\left( - \eta_{\gamma_i} P^{{ \delta}} \right)
\end{equation}
where $\eta_{\gamma_i}$ is a positive constant which is
independent of $P$ and $\hat{V}_i$ is the estimate for $V_i$ obtained by choosing the closest point in the constellation based on observation $Y_i$.

By Fano's inequality and the Markov chain $V_i \rightarrow Y_i\rightarrow\hat{V}_i$, we know that,
\begin{align}
  I(V_i; Y_i) & \ge I(V_i; \hat V_i) \\
  &  = H(V_i) - H(V_i|\hat V_i) \\
  & = \log( |\mathcal{V}_i| )- H(V_i|\hat V_i)\\
  & \ge \log( |\mathcal{V}_i| )- 1 - \mathrm{Pr}(V_i\neq \hat V_i)  \log ( |\mathcal{V}_i| ) \\
  & = \Big[1 -  \mathrm{Pr}(V_i\neq \hat V_i) \Big]\log( |\mathcal{V}_i| )- 1 \\
   & = \log( |\mathcal{V}_i| )- o(\log P) \label{eqn:kic:apply_lemma_2}\\
& =
\frac{ (K-1)M  (1-\delta)} { M_S + \delta }
\left(\frac{1}{2} \log P\right) + o(\log P)
\label{kic-ivy_low}
\end{align}
where $o(\cdot)$ is the little-$o$ function, $\mathcal{V}_i$ is the alphabet of $V_i$ and, in this case, the cardinality of $\mathcal{V}_i$ is $(2Q+1)^{(K-1) M} = (2Q+1)^{(K-1) m^{K(K-1)+2}}$. Here,  $M$ is defined in \eqref{eq:cardinality_real_ic}.

Now, we bound the second term in \eqref{eq:rate_ic_equation}. Let
\begin{align}
U \stackrel{\Delta}{=} \left\lbrace \mathbf{u}_i, \tilde{\mathbf{u}}_i, i=1,\ldots,K \right\rbrace
\end{align}
We have,
\begin{align}
I(V_i;Z|V_{-i}) =& I(V_i,U;Z|V_{-i}) - I(U;Z|V_1^K)\\
=& h(Z) - h(Z|U,V_1^K) - H(U|V_1^K) + H(U|Z,V_1^K)\\
\leq& \frac{1}{2}\log P - h(N_Z) - H(U) +o(\log P)\label{eq:equivocation_ic_step1}\\
=& \frac{1}{2}\log P - H(U) + o(\log P)\\
=& \frac{1}{2}\log P - \log (2Q+1)^{2KM}+o(\log P)\\
=& \frac{1}{2}\log P - \frac{(1-\delta)2KM}{2(M_S+\delta)}\log P + o(\log P)\label{eq:ic_eve_leakage}
\end{align}
Now, combining \eqref{kic-ivy_low} and \eqref{eq:ic_eve_leakage}, we have,
\begin{align}
R_i \geq \frac{2Km^{K(K-1)+2} - (K+1)(m+1)^{K(K-1)+2} -
M\delta (3K-1)}{(K-1)m^{K(K-1)}+(K+1)(m+1)^{K(K-1)+2}} \left(\frac{1}{2}\log P \right) + o(\log P)
\end{align}
By choosing $\delta$ small enough and choosing $m$ large enough, we can make $R_i$ arbitrarily close to $\frac{K-1}{2K}$. Thus, the sum s.d.o.f.~of $\frac{K-1}{2}$ is achievable with fixed channel gains.

\subsection{Fading Channel Gains}

Here, we present a scheme that achieves $\frac{K-1}{2}$ s.d.o.f.~using asymptotic vector space alignment with channel extension. Let $\Gamma = (K-1)^2$. We use $M_n = (K-1)n^\Gamma + (K+1)(n+1)^\Gamma$ channel uses to transmit $K(K-1)n^\Gamma$ message symbols securely to the legitimate receivers in the presence of the eavesdropper. Thus, we achieve a sum s.d.o.f.~of $\frac{K(K-1)n^\Gamma}{(K-1)n^\Gamma + (K+1)(n+1)^\Gamma}$, which gets arbitrarily close to $\frac{K-1}{2}$ as $n \rightarrow \infty$.

First, we divide each message into many sub-messages; specifically, the message of the $i$th transmitter, $W_i$, is divided into $(K-1)$ sub-messages $V_{ij}, j=1,\ldots,K+1, j\neq i,i+1$. Each $V_{ij}$ is encoded into $n^{\Gamma}$ independent streams $v_{ij}(1),\ldots,v_{ij}(n^\Gamma)$, which we denote as $\mathbf{v}_{ij} \stackrel{\Delta}{=} \left( v_{ij}(1),\ldots,v_{ij}(n^\Gamma)\right)^T $. We also require artificial noise symbols $U_i$ and $\tilde{U}_i$ at each transmitter $i$. Again, we encode the artificial noise symbols $U_i$ and $\tilde{U}_i$ as
\begin{align}
\mathbf{u}_i \stackrel{\Delta}{=}& \left( u_i(1),\ldots, u_i((n+1)^\Gamma)\right)^T, i=1,\ldots, K\\
\tilde{\mathbf{u}}_i \stackrel{\Delta}{=}& \left(\tilde{u}_i(1),\ldots,\tilde{u}_i(n^\Gamma)\right)^T, i=1,\ldots,K-1\\
\tilde{\mathbf{u}}_K \stackrel{\Delta}{=}& \left(\tilde{u}_i(1),\ldots,\tilde{u}_i((n+1)^\Gamma)\right)^T
\end{align}
In each channel use $t\leq M_n$, we choose precoding column vectors $\mathbf{p}_{ij}(t)$, $\mathbf{q}_i(t)$ and $\tilde{\mathbf{q}}_i(t)$  with the same number of elements as $\mathbf{v}_{ij}$, $\mathbf{u}_i$ and $\tilde{\mathbf{u}}_i$, respectively. In channel use $t$, transmitter $i$ sends
\begin{align}
X_i(t) = \sum\limits_{j} \mathbf{p}_{ij}(t)^T\mathbf{v}_{ij} + \mathbf{q}_i(t)^T\mathbf{u}_i + \tilde{\mathbf{q}}_i(t)^T\tilde{\mathbf{u}}_i
\end{align}
where we have dropped the limits on $j$ in the summation for notational simplicity.
By stacking the precoding vectors for all $M_n$ channel uses, we let,
\begin{align}
\mathbf{P}_{ij} = \left(\begin{array}{c}
\mathbf{p}_{ij}(1)^T\\
\vdots\\
\mathbf{p}_{ij}^T(M_n)
\end{array} \right), \qquad \mathbf{Q}_{i} = \left(\begin{array}{c}
\mathbf{q}_{i}(1)^T\\
\vdots\\
\mathbf{q}_{i}(M_n)^T
\end{array} \right), \qquad \tilde{\mathbf{Q}}_{i} = \left(\begin{array}{c}
\tilde{\mathbf{q}}_{i}(1)^T\\
\vdots\\
\tilde{\mathbf{q}}_{i}(M_n)^T
\end{array} \right)
\end{align}
Now, letting $\mathbf{X}_i = \left(X_i(1),\ldots,X_i(M_n) \right)^T $, the channel input for all transmitter $i$ over $M_n$ channel uses can be compactly represented as
\begin{align}
\mathbf{X}_i = \sum\limits_j \mathbf{P}_{ij}\mathbf{v}_{ij} + \mathbf{Q}_i\mathbf{u}_i + \tilde{\mathbf{Q}}_i\tilde{\mathbf{u}}_i
\end{align}

Recall that, channel use $t$, the channel output at receiver $l$ and the eavesdropper are, respectively, given by
\begin{align}
Y_l(t) =& \sum\limits_{k=1}^{K} h_{kl}(t)X_k(t) + N_l(t) \\
Z(t) =& \sum\limits_{k=1}^{K} g_{k}(t)X_k(t) + N_{Z}(t)
\end{align}
Let $\mathbf{H}_{kl} \stackrel{\Delta}{=} \mbox{diag}\left(h_{kl}(1),\ldots,h_{kl}(M_n) \right)$. Similarly, define $\mathbf{G}_{k} = \mbox{diag}\left(g_{k}(1),\ldots,g_{k}(M_n)\right) $. The channel outputs at receiver $l$ and the eavesdropper over all $M_n$ channel uses, $\mathbf{Y}_l = (Y_l(1),\ldots,\\Y_l(M_n) )^T$ and $\mathbf{Z} = \left(Z(1),\ldots,Z(M_n) \right)^T$, respectively, can be represented by
\begin{align}
\mathbf{Y}_l =& \sum\limits_{k=1}^{K} \mathbf{H}_{kl}\mathbf{X}_{k} + \mathbf{N}_l\\
=& \sum\limits_{k=1}^K \mathbf{H}_{kl}\left( \sum\limits_{\substack{j=1\\j\neq k,k+1}}^{K+1} \mathbf{P}_{kj}\mathbf{v}_{kj} + \mathbf{Q}_k\mathbf{u}_k + \tilde{\mathbf{Q}}_k\tilde{\mathbf{u}}_k \right)+ \mathbf{N}_l\\
=&\sum\limits_{\substack{j=1\\j\neq l,l+1}}^{K+1} \mathbf{H}_{ll}\mathbf{P}_{lj}\mathbf{v}_{lj} + \sum\limits_{\substack{k=1\\k\neq l}}^{K} \sum\limits_{\substack{j=1\\j\neq k,k+1}}^{K+1} \mathbf{H}_{kl}\mathbf{P}_{kj}\mathbf{v}_{kj} + \sum\limits_{k=1}^{K} \mathbf{H}_{kl}\left( \mathbf{Q}_k\mathbf{u}_k + \tilde{\mathbf{Q}}_k\tilde{\mathbf{u}}_k\right)+ \mathbf{N}_l
\label{eq:legit_sig}\\
\intertext{and,}
\mathbf{Z} =& \sum\limits_{k=1}^{K} \mathbf{G}_{k}\mathbf{X}_{k}+ \mathbf{N}_Z\\
=& \sum\limits_{k=1}^{K}\sum\limits_{\substack{j=1\\j\neq k,k+1}}^{K+1} \mathbf{G}_k\mathbf{P}_{kj}\mathbf{v}_{kj} + \sum\limits_{k=1}^{K} \mathbf{G}_{k}\left( \mathbf{Q}_k\mathbf{u}_k + \tilde{\mathbf{Q}}_k\tilde{\mathbf{u}}_k\right)+\mathbf{N}_Z\label{eq:eve_sig}
\end{align}

Note that receiver $l$ wants to decode $\mathbf{v}_{lj}, j=1,\ldots,K+1, j\neq l,l+1$. Thus, the remaining terms in \eqref{eq:legit_sig} constitute interference at the $l$th receiver. Recall that $CS(\mathbf{X})$ denotes the column space of the matrix $\mathbf{X}$.  Then, $I_l$ denoting the space spanned by this interference is
\begin{align}
I_l = \left( \bigcup\limits_{k\neq l,j\neq k,k+1} CS\left( \mathbf{H}_{kl}\mathbf{P}_{kj}\right)\right)   \bigcup \left(\bigcup\limits_{k=1}^K CS\left( \mathbf{H}_{kl}\mathbf{Q}_k\right)   \right) \bigcup \left(\bigcup\limits_{k=1}^K CS\left( \mathbf{H}_{kl}\tilde{\mathbf{Q}}_k\right)   \right)
\end{align}
Note that there are $(K-1)n^\Gamma$ symbols to be decoded by each legitimate receiver in $(K-1)n^\Gamma + (K+1)(n+1)^\Gamma$ channel uses. Thus, for decodability, the interference can occupy a subspace of rank at most $(K+1)(n+1)^\Gamma$, that is,
\begin{align}
\mbox{rank}(I_l) \leq (K+1)(n+1)^\Gamma
\end{align}
To that end, we align the noise and message subspaces at each legitimate receiver appropriately. Note that no such alignment is possible at the external eavesdropper since the transmitters do not have its CSI. However, note that we have a total of $(K-1)n^\Gamma + (K+1)(n+1)^\Gamma$ artificial noise symbols which will span the full received signal space at the eavesdropper and secures all the messages.

Fig.~\ref{fig:alignment_inter} shows the alignment for $K=3$ receivers. For the general $K$-user case, Fig.~\ref{fig:alignment_k_users} shows the alignment in the interfering signal dimensions. At receiver $l$, it is as follows:
First, the artificial noise symbols $\tilde{\mathbf{u}}_k$ is aligned with $\mathbf{u}_{k+1}$, for every $k=1,\ldots,K-1$.
Thus, we have,
\begin{align}
\mathbf{H}_{kl}\tilde{\mathbf{Q}}_{k} \preceq \mathbf{H}_{(k+1)l}\mathbf{Q}_{(k+1)},\quad k=1,\ldots,K-1
\end{align}
where $\mathbf{A} \preceq \mathbf{B}$ is used to denote that $CS(\mathbf{A}) \subseteq CS(\mathbf{B})$. Thus, the subspace spanned by the artificial noise symbols can have a rank of at most $(K+1)(n+1)^\Gamma$.

The unwanted message symbols $\mathbf{v}_{kj}$, $k\neq l$, are aligned with $\mathbf{u}_{j}$ if $j\leq K$, or $\tilde{\mathbf{u}}_K$ otherwise. Thus,
\begin{align}
\mathbf{H}_{kl}\mathbf{P}_{kj} &\preceq  \mathbf{H}_{jl}\mathbf{Q}_{j},\quad j\leq K\\
\mathbf{H}_{kl}\mathbf{P}_{k(K+1)} &\preceq \mathbf{H}_{Kl}\tilde{\mathbf{Q}}_{K}
\end{align}
for each $k\neq l$. Since, the unwanted messages at each receiver are aligned under the artificial noise subspaces, they do not increase the rank of $I_l$ any further.

We can group the alignment equations for the artificial noise $\mathbf{u}_k$, $k=1,\ldots,K$, and $\tilde{\mathbf{u}}_{K}$ for all $K$ legitimate receivers. For $\mathbf{u}_1$, we have,
\begin{align}
\mathbf{H}_{kl}\mathbf{P}_{k1} \preceq  \mathbf{H}_{1l}\mathbf{Q}_{1},\quad & k\in \left\lbrace 2,\ldots,K\right\rbrace, l\in \left\lbrace 1,\ldots,K\right\rbrace , l\neq k \label{eq:align1}
\end{align}
Clearly, these are $(K-1)^2$ alignment equations. Similarly, we have $(K-1)^2$ alignment equations for $\tilde{\mathbf{u}}_K$, given by
\begin{align}
\mathbf{H}_{kl}\mathbf{P}_{k(K+1)} \preceq \mathbf{H}_{Kl}\tilde{\mathbf{Q}}_{K}, \quad & k\in \left\lbrace 1,\ldots,K-1\right\rbrace, l\in \left\lbrace 1,\ldots,K\right\rbrace, l\neq k \label{eq:align_k_1}
\end{align}
For the artificial noises $\mathbf{u}_k, k=2,\ldots,K$, we have the following alignment equations:
\begin{align}
\mathbf{H}_{(k-1)l}\tilde{\mathbf{Q}}_{k-1} &\preceq \mathbf{H}_{kl}\mathbf{Q}_{k}\label{eq:align_k1}\\
\mathbf{H}_{il}\mathbf{P}_{ik} &\preceq  \mathbf{H}_{kl}\mathbf{Q}_{k}, \quad i\neq k-1,k,\quad
l\neq i \label{eq:align_k2}
\end{align}
Thus, there are $(K-1)^2+1$ alignment equations for each $\mathbf{u}_k, k=2,\ldots,K$. Now we make the following selections:
\begin{align}
\mathbf{P}_{k1} =& \tilde{\mathbf{P}}_1,\quad k=2,\ldots,K\\
\mathbf{P}_{k(K+1)} =& \tilde{\mathbf{P}}_{K+1}, \quad k=1,\ldots,K-1\\
\mathbf{P}_{ik} =& \tilde{\mathbf{P}}_{k}, \quad i\neq k-1,k,\quad k=2,\ldots,K\\
\mathbf{H}_{(k-1)1}\tilde{\mathbf{Q}}_{k-1} =& \mathbf{H}_{(k+1)1}\tilde{\mathbf{P}}_{k}, \quad k=2,\ldots,K-1\\
\mathbf{H}_{(K-1)2}\tilde{\mathbf{Q}}_{K-1} =& \mathbf{H}_{12}\tilde{\mathbf{P}}_{K}
\end{align}
Now, note that it suffices to choose the matrices $\tilde{\mathbf{P}}_{k}$, $k=1,\ldots,K+1$ in order to specify all the precoding matrices. Using these selections in our alignment equations in \eqref{eq:align1}, \eqref{eq:align_k_1}, \eqref{eq:align_k1} and \eqref{eq:align_k2}, we have $(K-1)^2$ alignment equations for each $\mathbf{u}_k$, $k=1,\ldots,K$ and $\tilde{\mathbf{u}}_K$, given by,
\begin{align}
\mathbf{T}_k \tilde{\mathbf{P}}_k \preceq& \mathbf{Q}_k, \quad \mathbf{T}_k \in \tau_k,\quad k=1,\ldots,K\\
\mathbf{T}_{K+1}\tilde{\mathbf{P}}_{K+1} \preceq& \tilde{\mathbf{Q}}_K,\quad \mathbf{T}_{K+1} \in \tau_{K+1}
\end{align}
where the sets $\tau_k$, $k=1,\ldots,K+1$ are given by
\begin{align}
\tau_1 =& \left\lbrace \mathbf{H}_{1l}^{-1}\mathbf{H}_{kl}, k\in \left\lbrace 2,\ldots,K\right\rbrace, l\in \left\lbrace 1,\ldots,K\right\rbrace, l\neq k  \right\rbrace \\
\tau_{K+1} =& \left\lbrace \mathbf{H}_{Kl}^{-1}\mathbf{H}_{kl}, k\in \left\lbrace 1,\ldots,K-1\right\rbrace, l\in \left\lbrace 1,\ldots,K\right\rbrace, l\neq k \right\rbrace\\
\tau_k =& \tau_k^{P} \bigcup \tau_k^Q
\end{align}
where,
\begin{align}
\tau_k^P =& \left\lbrace \mathbf{H}_{kl}^{-1}\mathbf{H}_{il},  i\notin \left\lbrace k-1,k\right\rbrace ,
l\neq i, l \in \left\lbrace 1,\ldots,K \right\rbrace  \right\rbrace \\
\tau_k^Q =& \begin{cases} \left\lbrace \mathbf{H}_{kl}^{-1}\mathbf{H}_{(k-1)l}\mathbf{H}_{(k-1)1}^{-1}\mathbf{H}_{(k+1)1}, l\in \left\lbrace 1,\ldots,K \right\rbrace  \right\rbrace, \quad \mbox{if } k\in \left\lbrace 2,\ldots,K-1 \right\rbrace  \\
\left\lbrace \mathbf{H}_{Kl}^{-1}\mathbf{H}_{(K-1)l}\mathbf{H}_{(K-1)2}^{-1}\mathbf{H}_{12}, l\in \left\lbrace 1,\ldots,K \right\rbrace  \right\rbrace,\quad \mbox{if } k=K
\end{cases}
\end{align}
We can now construct the matrices $\tilde{\mathbf{P}}_k, k=1,\ldots,K+1$, $\mathbf{Q}_k, k=1,\ldots,K$ and $\tilde{\mathbf{Q}}_{K}$ as in \cite{cadambe_jafar_interference}
\begin{align}
\tilde{\mathbf{P}}_k =& \left\lbrace \left( \prod\limits_{\mathbf{T}\in \tau_k} \mathbf{T}^{\alpha_{T}}\right)\mathbf{w}_k : \alpha_{T} \in \left\lbrace 1,\ldots,n\right\rbrace   \right\rbrace \\
{\mathbf{Q}}_k =& \left\lbrace \left( \prod\limits_{\mathbf{T}\in \tau_k} \mathbf{T}^{\alpha_{T}}\right)\mathbf{w}_k : \alpha_{T} \in \left\lbrace 1,\ldots,n+1\right\rbrace   \right\rbrace\\
\tilde{\mathbf{Q}}_K =& \left\lbrace \left( \prod\limits_{\mathbf{T}_\in \tau_{K+1}} \mathbf{T}^{\alpha_{T}}\right)\mathbf{w}_{K+1} : \alpha_{T} \in \left\lbrace 1,\ldots,n+1\right\rbrace   \right\rbrace
\end{align}
where each $\mathbf{w}_k$ is the $M_n\times 1$ column vector containing elements drawn independently from a continuous distribution with bounded support. This completes the description of our scheme.

\textbf{Decodability:} By our construction, the interference space at legitimate receiver $l$ is given by,
\begin{align}
I_l= \left( \bigcup_{k=1}^{K} CS(\mathbf{H}_{kl}\mathbf{Q}_k)\right)\bigcup\left(CS(\mathbf{H}_{Kl}\tilde{\mathbf{Q}}_K) \right)
\end{align}
and clearly,
\begin{align}
\mbox{rank}(I_l) \leq (K+1)(n+1)^\Gamma
\end{align}
We only need to show that desired signals $\mathbf{v}_{lj}, j\neq l,l+1$ fall outside $I_l$. The desired signal space at receiver $l$ is given by
\begin{align}
\mathbf{D}_l = \left[ \mathbf{H}_{ll}\tilde{\mathbf{P}}_1\ldots \mathbf{H}_{ll}\tilde{\mathbf{P}}_{l-1}\quad \mathbf{H}_{ll}\tilde{\mathbf{P}}_{l+2}\ldots,\mathbf{H}_{ll}\tilde{\mathbf{P}}_K   \right]
\end{align}
We want to show that the matrix
\begin{align}
\mathbf{\Lambda}_l = \left[\mathbf{D}_l\quad \tilde{\mathbf{I}}_l\right]
\end{align}
where,
\begin{align}
\tilde{\mathbf{I}}_l=\left[  \mathbf{H}_{1l}\mathbf{Q}_1\ldots \mathbf{H}_{Kl}\mathbf{Q}_K\quad \mathbf{H}_{Kl}\tilde{\mathbf{Q}}_K\right]
\end{align}
is full rank almost surely. To do so, we will use \cite[Lemmas 1, 2]{cadambe_jafar_x_channel}. Note that the $m$th row of $\mathbf{H}_{kl}\mathbf{Q}_k$ contains the term $w_{mk}$ with exponent $1$, but no $w_{mk'}$ for $k\neq k'$, where $w_{mk}$ denotes the element in the $m$th row of $\mathbf{w}_k$. In fact, the term $w_{mk}$ occurs nowhere else in the matrix $\mathbf{\Lambda}_l$ except in $\mathbf{H}_{kl}\mathbf{Q}_k$ and $\mathbf{H}_{ll}\tilde{\mathbf{P}}_k$. This shows, using \cite[Lemmas 1, 2]{cadambe_jafar_x_channel}, that $\mathbf{D}_l$ and $\tilde{\mathbf{I}}_l$ are full rank almost surely. Further, it suffices to show that the matrices $\left[ \mathbf{H}_{ll}\tilde{\mathbf{P}}_k\quad \mathbf{H}_{kl}\mathbf{Q}_k \right], k=1,\ldots,K$, and $\left[ \mathbf{H}_{ll}\tilde{\mathbf{P}}_{K+1}\quad \mathbf{H}_{Kl}\tilde{\mathbf{Q}}_K \right]$ are all full column rank. First, $\left[ \mathbf{H}_{ll}\tilde{\mathbf{P}}_1\quad \mathbf{H}_{kl}\mathbf{Q}_1 \right]$ is full column rank since $\mathbf{H}_{kl}\mathbf{Q}_1$ misses the term $\mathbf{H}_{ll}$. Similarly, $\left[ \mathbf{H}_{ll}\tilde{\mathbf{P}}_{K+1}\quad \mathbf{H}_{Kl}\tilde{\mathbf{Q}}_1 \right]$ is full column rank. Further, if $k\neq l,l+1$, $\mathbf{H}_{kl}\mathbf{Q}_k$ does not contain $\mathbf{H}_{ll}$ and hence $\left[ \mathbf{H}_{ll}\tilde{\mathbf{P}}_k\quad \mathbf{H}_{kl}\mathbf{Q}_k \right]$ is full column rank. Finally, note that the $l$th transmitter does not transmit any message signals along $\tilde{\mathbf{P}}_k$, when $k=l,l+1$. Thus, the matrix $\mathbf{\Lambda}_l$ is full rank almost surely. This ensures decodability of the desired signals at each receiver.

\textbf{Security guarantee:} Let $\mathbf{v} =\left\lbrace \mathbf{v}_{ij}, i,j \in \left\lbrace 1,\ldots,K\right\rbrace ,j\neq i,i+1 \right\rbrace$, that is, $\mathbf{v}$ is the collection of all legitimate messages to be secured from the eavesdropper. Also, let $\mathbf{u} = \left\lbrace \mathbf{u}_k, \tilde{\mathbf{u}}_k, k=1,\ldots,K \right\rbrace $, that is $\mathbf{u}$ is the collection of all the artificial noise symbols. We note that \begin{align}
I(\mathbf{v} ;\mathbf{Z}) =& h(\mathbf{Z}) - h(\mathbf{Z}|\mathbf{v})\\
\leq& \frac{M_n}{2}\log P - h(\mathbf{A}\mathbf{u})+o(\log P)\\
=& \frac{M_n}{2}\log P - \frac{M_n}{2}\log P +o(\log P)\label{eq:step_inter1} \\
=& o(\log P)
\end{align}
where $\mathbf{A}$ is a $M_n\times M_n$ full rank matrix, and we have used Lemma~\ref{lem:dif_entropy} in \eqref{eq:step_inter1}. Also, we have implicitly used the fact that $\Omega$ appears in the conditioning of each mutual information and differential entropy term in the above calculation. Now, as before, by treating the vector channel with $M_n$ slots as one channel use, and using wiretap channel codes, we get,
\begin{align}
R_i \geq \frac{(K-1)n^{\Gamma}}{M_n} \log P + o(\log P)
\end{align}
for each $i=1,\ldots,K$, which gives us the required sum s.d.o.f.~of $\frac{K(K-1)n^{\Gamma}}{(K-1)n^\Gamma+(K+1)(n+1)^{\Gamma}}$, which approaches $\frac{K-1}{2}$ as $n\rightarrow \infty$.

\end{appendices}

\bibliographystyle{unsrt}
\bibliography{references}

\end{document}